\documentclass[fleqn,usenatbib]{mnras}
\usepackage{newtxtext,newtxmath, import}

\usepackage{graphicx, amsmath, bbm, mathtools, capt-of, orcidlink} 
\usepackage{stfloats}
\usepackage{siunitx}
\usepackage{footmisc}
\usepackage{float}
\usepackage{diagbox}
\usepackage[para]{threeparttablex}
\usepackage{eqnarray, physics}

\usepackage[T1]{fontenc}
\usepackage[dvipsnames]{xcolor}
\usepackage{amssymb}
\usepackage{pifont}
\newcommand{\cmark}{\ding{51}}%
\newcommand{\xmark}{\ding{55}}%
\usepackage[switch]{lineno}
\DeclareSIUnit\angstrom{\text {Å}}

\title[First measurement of narrow-line flux ratios for a lensed quasar with JWST/NIRSpec IFS.]{First measurement of narrow-line flux ratios for a lensed quasar with JWST/NIRSpec IFS.}

\author[H. Paugnat et al.]{Hadrien Paugnat \orcidlink{0000-0002-2603-6031}$^{1}$\thanks{E-mail: hpaugnat@astro.ucla.edu},
Tommaso Treu \orcidlink{0000-0002-8460-0390},$^{1}$,
Anna M. Nierenberg \orcidlink{0000-0001-6809-2536}$^{2}$,
Anowar J. Shajib \orcidlink{0000-0002-5558-888X}$^{3,4,5}$,
\newauthor
Shawn Knabel \orcidlink{0000-0001-5110-6241}$^{1}$, and
Daniel Gilman \orcidlink{0000-0002-5116-7287}$^{3}$\thanks{Brinson Prize Fellow}.
\\
$^{1}$Department of Physics and Astronomy, UCLA, Los Angeles, CA 90095-1547, USA\\
$^{2}$University of California, Merced, 5200 N Lake Road, Merced, CA 95341, USA\\
$^{3}$Department of Astronomy \& Astrophysics, University of Chicago, Chicago, IL 60637, USA\\
$^{4}$Kavli Institute for Cosmological Physics, University of Chicago, Chicago, IL 60637, USA\\
$^{5}$Center for Astronomy, Space Science and Astrophysics, Independent University, Bangladesh, Dhaka 1229, Bangladesh\\
\vspace{0.3cm}
}

\date{\vspace{-1.5cm}}
\pubyear{\the\year{}}

\begin{document}
\label{firstpage}
\pagerange{\pageref{firstpage}--\pageref{lastpage}}
\maketitle

\begin{abstract}
To determine the nature of dark matter (DM) using astrophysical probes, the next milestone is to constrain the properties of
structure on subgalactic scales. Gravitational lensing offers a remarkably powerful probe of this regime: in particular, statistics of flux-ratio anomalies (discrepancies between mass model predictions and observed flux ratios) in quadruply imaged quasars are sensitive to perturbations by low-mass DM halos down to $\sim 10^6 M_\odot$, independent of their baryonic content. Studies leveraging these anomalies require high-quality flux-ratio measurements from an emission region insensitive to stellar microlensing. In this paper, we present the first measurement of narrow-line flux ratios for a gravitationally lensed quasar using JWST/NIRSpec with Integral Field Spectroscopy (IFS), targeting the well-studied system RXJ1131$-$1231. Flux ratios are extracted from the [\ion{S}{III}] $9071/9533~\si{\angstrom}$ narrow-line doublet — the first use of this doublet for substructure studies — by performing a full lens model reconstruction to isolate the unresolved nuclear emission from extended narrow-line emission. The resulting spectra are jointly modeled using \texttt{lensqso-specfit}, a new, publicly available software package introduced in this work for the simultaneous spectral fitting of multiple lensed quasar images.
We achieve $\sim$5\% uncertainties on the flux ratios, comparable to the precision of JWST/MIRI warm dust measurements, and detect a clear anomaly in the cusp images relative to a standard smooth lens model. Our results are in good agreement with previous narrow-line measurements and broadly consistent with JWST/MIRI warm dust flux ratios, with marginal ($\sim 2-3$ $\sigma$) deviations. We illustrate how such systematic shifts between differently sized emission regions may be enhanced by small spatial offsets of the order of tens of parsecs in the presence of a low-mass dark matter halo. Our method is generalizable to other systems with existing or future IFS observations, and the combination of narrow-line and warm dust flux ratios offers a new avenue for improving DM substructure constraints with flux-ratio anomaly statistics.
\vspace{0.2cm}
\end{abstract}

\begin{keywords}
gravitational lensing: strong--cosmology: dark matter--quasars: emission lines
\vspace{-0.2cm}
\end{keywords}

\section{Introduction} \label{sec:intro}

The Cold Dark Matter (CDM) paradigm, embedded within the standard $\Lambda$CDM cosmological model, has proven remarkably successful at explaining the large-scale structure of the Universe. On mass scales $\gtrsim10^{11}M_\odot$, predictions from CDM match observations of the Cosmic Microwave Background \citep[e.g.,][]{Planck2020}, of the matter power spectrum with galaxy surveys \citep[e.g.,][]{Tegmark2004}, and of the density profiles of galaxy-hosting DM halos \citep[e.g.,][]{Weinberg2015}. 

The frontier has therefore shifted to subgalactic scales (low-mass halos, subhalos, and central regions of massive halos), where the matter power spectrum is poorly constrained. The CDM paradigm has been facing a number of persistent challenges on these scales \citep[for a review, see][]{Bullock2017}, prompting the development of many alternative DM models—for instance, warm DM \citep[e.g.,][]{Bode2001}, wave DM \citep[e.g.,][]{Hui2021}, and self-interacting DM \citep[e.g.,][]{TulinYu2018}—which predict widely different subhalo abundances and density profiles. It remains debated, however, whether these tensions signal a breakdown of CDM or can be explained by complex baryonic physics \citep[e.g.,][]{Wetzel2016}. This motivates the need for robust observational constraints on subgalactic scales that are independent of baryonic processes, as these could provide critical insight on the nature of DM.

Strong gravitational lensing offers a powerful approach to this challenge: DM structures within strong gravitational lens galaxies and along their line of sight leave a gravitational imprint, regardless of their baryonic content. This is a crucial advantage over other methods such as dwarf galaxy surveys \citep[e.g.,][]{Nadler2024} and measurements of the Lyman-alpha forest power spectrum \citep[e.g.,][]{Villasenor2023, Irsic2024}: not only does gravitational lensing not require explicit modeling of baryonic physics, but it can also probe the halo mass function at lower mass scales, well below the threshold at which DM structures are expected to host stars \citep[e.g.,][]{Treu2010, Vegetti2024, Nierenberg2026}. 

\newpage
Direct detection of individual subhalos is possible from the perturbations that they impart on the imaging data of a given lensed system. This approach has enabled a small number of significant detections \citep[e.g.,][]{Vegetti2010, Vegetti2012, Hezaveh2016, Nightingale2024, Amvrosiadis2026}, as well as investigations of the inner structure of subhalos \citep[e.g.,][]{Minor2021a, Minor2021b, Despali2025}, with limitations due to the large computational cost and systematic errors in the lens mass model \citep[e.g.,][]{Nightingale2024, Stacey2025} and source light reconstruction  \citep[e.g.,][]{Ballard2024, Ephremidze2025}. It is mainly sensitive to the most massive perturbers \citep[ $\gtrsim 10^9 M_\odot$, ][]{Ritondale2019}, though high-resolution observations of lensed arcs in the radio have extended the method to lower masses \citep{Powell2025}.

Another productive and complementary technique exploits population-level statistics of flux ratios in quadruply imaged quasars (“quads”), which are sensitive to the collective effect of low-mass subhalos and line-of-sight halos down to $\sim 10^6-10^7\,M_\odot$ \citep[e.g.,][]{Nierenberg2017, Gilman2019}, to infer the properties of DM substructure. It leverages the fact that the point-source magnifications are sensitive to small scale perturbations in the lens potential caused by low-mass halos, while the image positions are mostly determined by the lens mass distribution on larger scales. Thus, early studies \citep{MaoSchneider1998, MetcalfMadau2001, MetcalfZhao2002} proposed that the presence of subhalos could explain the observed “flux-ratio anomalies” (discrepancies between the observed flux ratios of quasar image pairs and the relative magnifications expected from the smooth mass distribution) in some individual quad systems, with the first statistical tests of CDM following soon after \citep{DalalKochanek2002}. Since then, flux-ratio anomaly statistics have become one of the primary tools for constraining the particle nature of dark matter using strong lensing \citep{Gilman2019, Gilman2020, Hsueh2020, JWST_LQ_DM_II, JWST_LQ_DM_IV}, helped by major improvements in galaxy-scale lensing observations: extended lensed arcs from the quasar host galaxy can now be detected and leveraged (jointly with the quasar image positions) to provide better constraints on the smooth mass distribution \citep{Gilman2024, CAB_paper_I}.

A critical practical consideration in those studies is the physical size of the emission region around the quasar that is being used for flux-ratio measurements. In order to filter out the microlensing signal from stars in the foreground galaxy, one needs to focus on background sources that are significantly larger in projection than the Einstein radius of a star ($\sim 1 \mu$as at a typical lens redshift $z\sim0.5$). This leads to considering sources with angular sizes $\gtrsim 0.1$~mas ($\gtrsim$~0.1~pc across for a typical source redshift $z\sim2$). It also ensures that the light-crossing time exceeds the time delays  between quasar images \citep[of the order of days to months,][]{Birrer2024}, such that the intrinsic variability of the source has a negligible impact on the flux ratios. At the same time, the emission region must remain compact enough to be sensitive to perturbations from dark matter subhalos (known as “millilensing” since it typically occurs on $\sim $ mas scales). Examples of targeted emission regions include the compact radio core \citep{Lee2017, Kim2022}, and the warm dust component which dominates the quasar spectral energy distribution in the rest-frame mid-infrared ($\sim 8-12 \mu$m). The warm dust emission, in particular, originates from a $\sim 0.1-10$ pc region \citep{Burtscher2013, Sluse2013} and is thus particularly sensitive to millilensing \citep[e.g.,][]{Chiba2005, MacLeod2009, MacLeod2013, Jones2019}. This has motivated a recent survey to measure warm dust flux ratios in $\sim30$ lensed quasars using multiband Mid-Infrared Instrument (MIRI) imaging with the James Webb Space Telescope (JWST) \citep{JWST_LQ_DM_I, JWST_LQ_DM_II, JWST_LQ_DM_III, JWST_LQ_DM_IV}.

The quasar narrow-line region (NLR) is another excellent target for flux-ratio studies \citep{MoustakasMetcalf2003}: typically dominated by an unresolved nuclear component with sizes of $\sim 10-100$~pc \citep{Schmitt2003b, Schmitt2003a, Muller-Sanchez2011, Nierenberg2014}, it is less compact than the warm dust emission region while still being sensitive to substructure, and therefore provides a viable alternative for flux-ratio anomaly statistics. Previous work has successfully exploited NLR emission (specifically, the narrow doublets [\ion{O}{III}] $4959/5007\ \si{\angstrom}$ and [Ne~III] $3870/3969\ \si{\angstrom}$), either observed with Hubble Space Telescope (HST) grism spectroscopy \citep{Nierenberg2017, Nierenberg2020} or with ground-based IFS using adaptive optics \citep{Nierenberg2014}. The integral field unit (IFU) mode of the Near-Infrared Spectrograph (NIRSpec) onboard JWST \citep{Boker2022} now offers a powerful new avenue for measuring narrow-line flux ratios with unprecedented sensitivity and precision.

In this work, we present the first measurement of narrow-line flux ratios for a lensed quasar using the JWST NIRSpec IFU. This is also the first flux-ratio measurement using the narrow doublet [\ion{S}{III}] 9071/9533 $\si{\angstrom}$. We choose the well-studied system RXJ1131$-$1231 \citep{Sluse2003}, and present a method that can be generalized to other quads that have been or will be observed with IFS. We report narrow-line flux ratio values with $\sim 5 \%$ uncertainties and compare them to the magnification ratios predicted by the smooth mass model, finding a significant flux-ratio anomaly between the cusp images. We compare our results to previous flux-ratio measurements, establishing an overall consistency, though we identify marginal differences ($\sim2-3~\sigma$) with respect to the values measured by the recent JWST/MIRI survey targeting warm dust emission \citep{JWST_LQ_DM_III}. Using a simple toy model, we show that small ($\sim 10$~pc) spatial offsets between these emission regions, in addition to their size difference, may contribute to systematic shifts in the flux ratios.

This paper is organized as follows. Section~\ref{sec:data} presents our observations and data-reduction method. Section~\ref{sec:method} explains how we obtained flux-ratio measurements, giving details regarding the extraction of the quasar images point-source spectra, and the spectral modeling. In Section~\ref{sec:results}, we present our final measurements and discuss these results in perspective with the literature. Finally, in Section~\ref{sec:conclusion}, we summarize our conclusions and discuss future prospects for this method. Throughout this paper, we assume a flat $\Lambda$CDM cosmology with $h = 0.7$ and $\Omega_m = 0.3$.

\vspace{-0.2cm}

\section{Observations \& data reduction}
\label{sec:data}

\begin{figure*}
    \centering
    \includegraphics[width=0.95\textwidth]{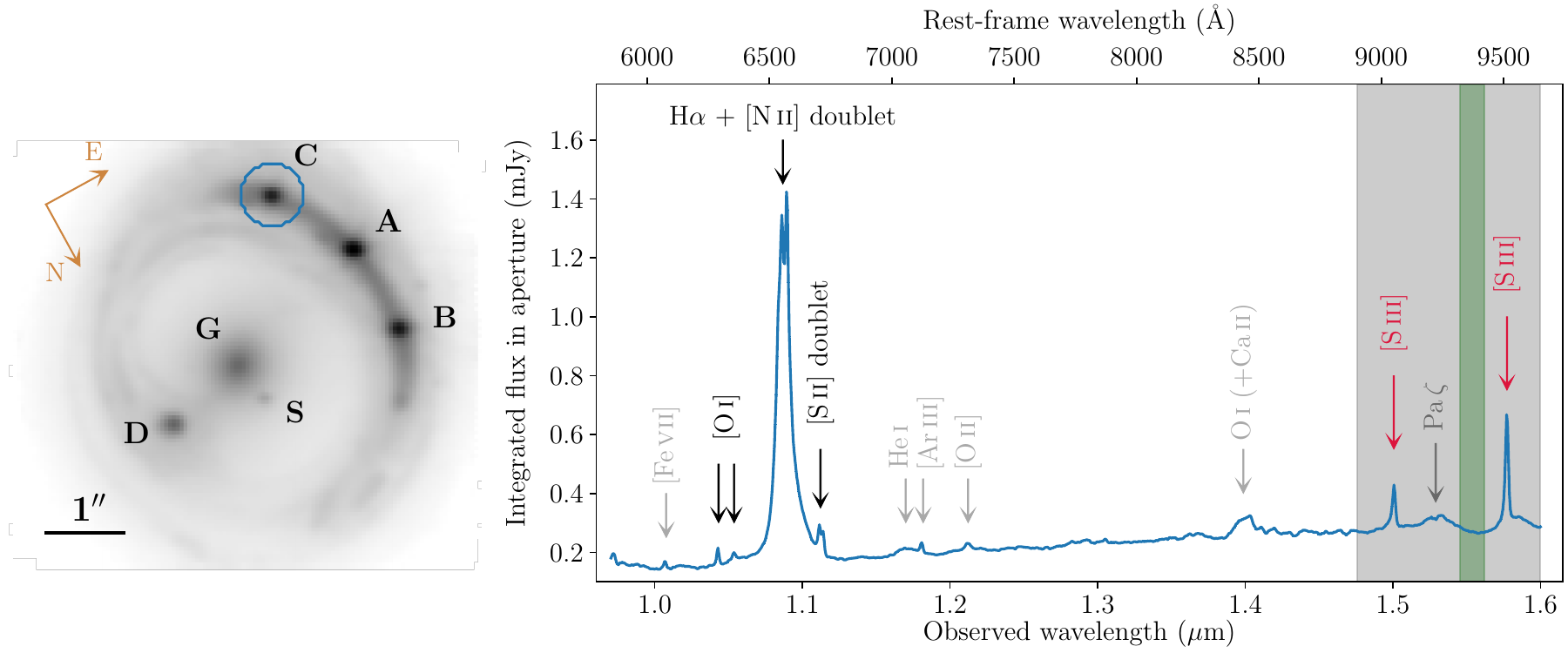}
    \caption{\textbf{Left panel:} “White-light” image of RXJ1131$-$1231 from the JWST/NIRSpec datacube, summed over the entire wavelength range. The four quasar images (A, B, C, and D) are labeled following the convention from previous studies of the same system \protect\citep{Sluse2007, Sugai2007, JWST_LQ_DM_III, JWST_LQ_DM_IV, Shajib2026}, for direct comparison of the flux ratios. The main deflector and its nearby satellite are marked with G and S, respectively. In blue, we show a circular aperture of radius $0\farcs4$ (8 pixels) around image C. \textbf{Right panel:} Spectrum integrated over that aperture, with lines identified using the AGN spectral atlas from \protect\cite{VandenBerk2001} for the rest-frame optical, and \protect\cite{Riffel2006} for the rest-frame near-infrared (NIR). }
    \label{fig:wavelength_ranges}
\end{figure*}

The lens system RXJ1131$-$1231 (J2000: ${\rm RA}=172.9644^{\circ}, {\rm DEC}=-12.5329^{\circ}$) is a quadruply imaged quasar, first discovered by \cite{Sluse2003}. The deflector is an elliptical galaxy at a relatively low redshift \citep[$z_l=0.295$,][]{Sluse2003}, making this system one of the brightest and best-resolved known quad lenses, and therefore, an ideal target for high signal-to-noise spectroscopic observations. The prominent lensed arcs display complex structure (see Figure~\ref{fig:wavelength_ranges}), which provide strong constraints on the lens mass model—a key ingredient for many strong lensing science cases.

Integral field spectroscopy of RXJ1131$-$1231 with JWST/NIRSpec was obtained as part of Cycle 1 program GO-1794 (PI: Suyu; co-PIs: Yıldırım, Treu), designed to provide state-of-the-art spatially resolved kinematic data for time-delay cosmography \citep{TDCOSMO2025,Shajib2026}. Observations were carried out on May 9, 2023 (UT) and followed a $2\times 2$ mosaic pattern \citep[see Figure~1 from][]{Shajib2026}, resulting in a $6\farcs1 \times 5\farcs55$ field of view over the system, large enough to cover the quasar images and the lensed arcs. The G140M grating and the F100LP filter were used, providing a nominal spectral resolution $R\sim 1000$ over the observed $0.97 \si{\micro\m} - 1.84\si{\micro\m}$ wavelength range. At the source redshift of $z_s=0.658$ \citep{Sluse2003, Sluse2007}, this range includes several forbidden doublets that make good candidates for narrow-line flux-ratio measurements (see Section~\ref{sec:method}): namely, the [\ion{O}{I}] 6300/6363 $\si{\angstrom}$, [\ion{N}{II}] 6548/6583 $\si{\angstrom}$, [\ion{S}{II}] 6716/6731 $\si{\angstrom}$, and [\ion{S}{III}] 9071/9533 $\si{\angstrom}$ doublets. The presence of these doublets, combined with the fact that all four quasar images are visible, make this dataset well-suited for the ancillary science case of narrow-line flux-ratio measurements.

Details about the observing strategy are as follows. For each tile in the mosaic, observations followed a 4-point dithering pattern with shifts $<1^{\prime\prime}$. Each dither position consisted of a 1050.4~s exposure, divided into two integrations and seven groups in order to prevent saturation in the bright quasar pixels. Calibration exposures using the same integration time were obtained separately, including a background exposure from a nearby empty patch of sky, and a multi-shutter array leakage exposure for each mosaic tile.

The data were reduced using the custom reduction pipeline $\texttt{RegalJumper}$\footnote{\url{https://github.com/ajshajib/regaljumper}}, which extends the standard JWST pipeline by adding cleaning steps for artifacts, cosmic rays, and $1/f$ noise \citep[for a detailed description, see][]{Shajib2026}. In particular, the datacube was built by drizzling onto a grid with a pixel size of $0\farcs05$. The reduced datacube still presented wave-like artifacts in individual spaxel spectra (also known as “wiggles”) due the undersampling of the point spread function (PSF) \citep[e.g.,][]{Law2023, Dumont2025}. \cite{Shajib2026} corrected this effect for the spaxels around the lensing galaxy (i.e., only those relevant for their science case) using the software  package \texttt{raccoon} \footnote{ \url{https://github.com/ajshajib/raccoon}}\citep{raccoon}—we perform a similar cleaning on the spaxels near the quasar images and in the brightest parts of the lensed arcs. The settings used for this procedure are detailed in Appendix~\ref{app:wiggle_correction}.

\section{Narrow-line flux-ratio measurement}
\label{sec:method}

In this section, we describe the analysis steps taken on the reduced datacube to extract the quantities that can be used for dark matter substructure inference, i.e., measurements of the ratios of nuclear narrow-line flux between quasar images, propagating the uncertainties associated with the point-source image positions, extended emission at the relevant wavelength, and PSF reconstruction. We first detail, in Section~\ref{subsec:quasar_spectra_extraction}, how we extract the spectra corresponding to the unresolved quasar component (removing the contribution from extended emission), reviewing our approach to estimate the PSF (Section~\ref{subsubsec:PSF_estimation}), construct a lens model (Section~\ref{subsubsec:lens_model}), then isolate the quasar contribution at each wavelength (Section~\ref{subsubsec:spectra_extraction}). In Section~\ref{subsec:spectral_model}, we explain how we modeled the four quasar spectra that were extracted from the datacube.

\subsection{Extraction of quasar spectra}
\label{subsec:quasar_spectra_extraction}

For the characterization of DM substructure, flux-ratio measurements can target a compact ($\lesssim$100 pc), unresolved nuclear region which is expected to dominate the narrow-line emission \citep[e.g.,][]{Muller-Sanchez2011, Nierenberg2014, Nierenberg2017}. High-resolution observations of luminous quasars at low redshifts ($z<1$), however, reveal that some narrow-line emission can also originate from a much more extended narrow-line region (ENLR) with low surface brightness, that can span up to $\sim$ kpc scales and display irregular morphologies \citep[e.g.,][]{Bennert2002, Bennert2006, Chen2019}. Due to its size, this ENLR is insensitive to millilensing, so any significant flux contribution has to be removed in order to avoid “diluting” the flux-ratio anomaly signal from DM substructure. 
Similarly, if the quasar host galaxy is bright at the observed narrow-line wavelengths, its flux contribution can bias the observed flux ratios toward less anomalous values. Therefore, if they are prominent, these components (which we regroup under the term of “extended emission”) must be removed for the final flux-ratio measurements.

In our reduced datacube for RXJ1131$-$1231, extended emission is seen at all the narrow-line wavelengths in the form of bright lensed arcs (see, e.g., Figures~\ref{fig:PSF_range_image_model} and \ref{fig:spectra_range_image_model}). The quasar spectra can therefore not be simply extracted with apertures or a point-source model ; instead, we generated a full lens model to explicitly account for extended emission and separate its contribution from the nuclear narrow-line component. A similar approach has previously been employed for flux-ratio measurements with HST grism data \citep{Nierenberg2017, Nierenberg2020}, and has been validated on simulated data for the Keck OH-Suppressing Infrared Integral Field Spectrograph (OSIRIS)  \citep{PerezMendoza2026}, showing that $<5\%$ accuracy on the flux ratios can be achieved even with very bright extended emission when explicitly modeled. We proceeded in three steps, described in detail in the following paragraphs.

\subsubsection{PSF estimation}
\label{subsubsec:PSF_estimation}

\begin{figure*}
    \centering
    \includegraphics[width=0.99\textwidth]{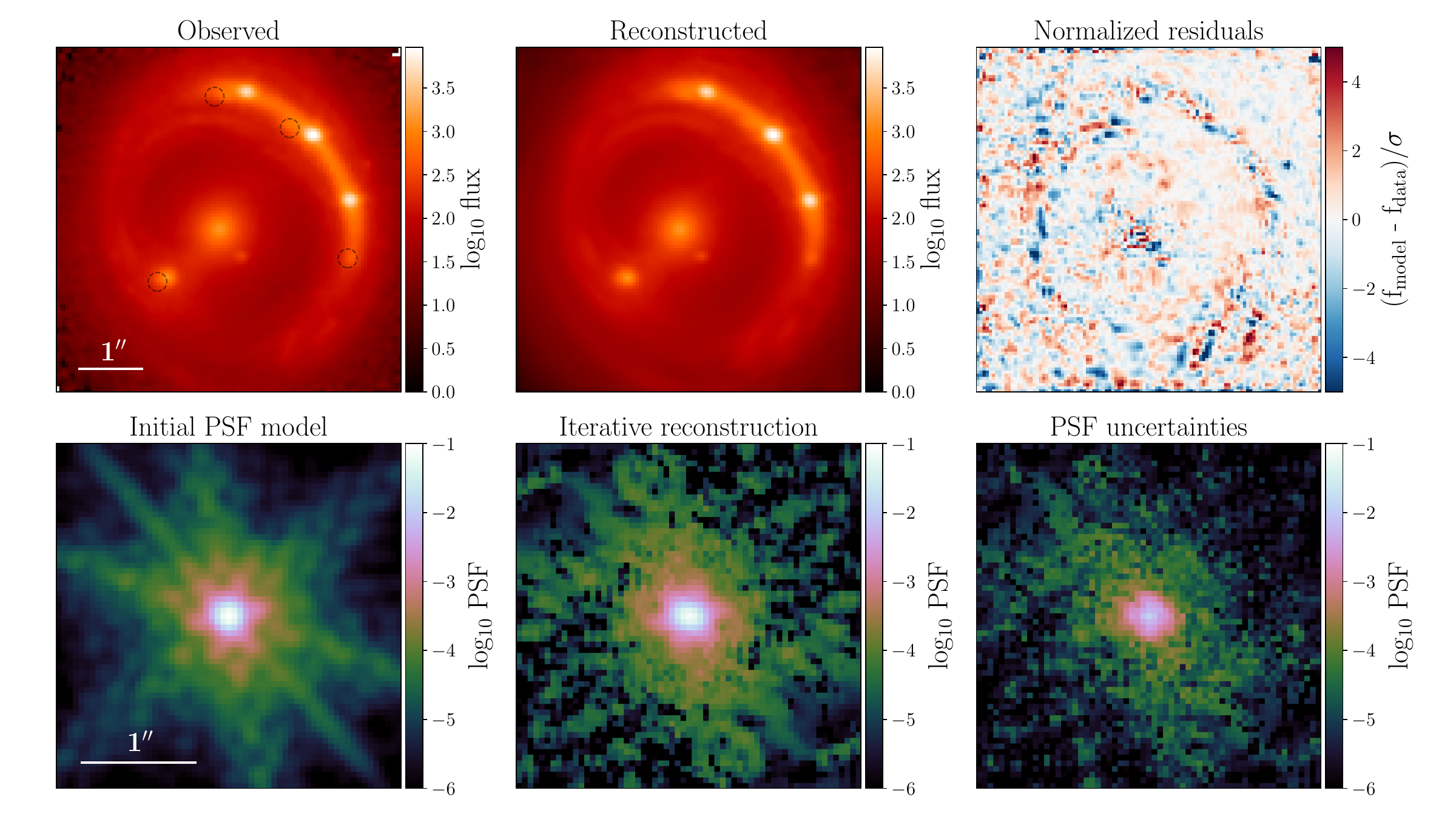}
    \caption{Image model for the PSF reconstruction step. \textbf{Top row:} (Left) White-light image summed over the $9320-9420\ \si{\angstrom}$ wavelength range (in green in Figure~\ref{fig:wavelength_ranges}). The dashed black circles show the location of a faint, compact light source, quadruply imaged but offset from the quasar images, that was explicitly included in the source light model (see Section~\ref{subsubsec:lens_model}). (Middle) Reconstructed light under the best-fit lens model. (Right) Residuals, normalized by the estimated noise map. \textbf{Bottom row,} from left to right: initial PSF estimate in that range generated with \texttt{stpsf}, output of the PSF reconstruction process, and map of total per-pixel uncertainties on the final PSF model (see Appendix~\ref{app:PSF_uncertainties}).}
    \label{fig:PSF_range_image_model}
    \vspace{-0.5cm}
\end{figure*}

In the first step, we characterized the PSF using a “white-light” image obtained by wavelength-collapsing the datacube in a small range (which improves the signal-to-noise ratio per spaxel compared to images at individual wavelengths). The range was selected to be (i) relatively small in order to avoid large changes in the light distribution within it, (ii) close to the targeted narrow-line emission feature(s) to safely neglect the wavelength dependence of the PSF, and (iii)  maximizing the contrast between the extended and unresolved emission, such that the four quasar images provide better point-source references for the PSF reconstruction. For the extraction of the quasar spectra around the [\ion{S}{III}] 9071/9533 $\si{\angstrom}$ doublet, we chose to model the PSF using the $9320-9420\ \si{\angstrom}$ range, shown in green in Figure~\ref{fig:wavelength_ranges} on top of a single-spaxel spectrum at the location of one of the quasar images. The corresponding white-light image is shown in the upper left panel of  Figure~\ref{fig:PSF_range_image_model}. A noise map for this image was obtained by taking the quadrature sum, in the relevant range, of the background noise level estimated from the separate background exposure, and the noise estimated by the reduction pipeline (which accounts for Poisson noise in the detector, read noise, and flat-field uncertainties).

This white-light image was fit using a standard lens modeling procedure, with the publicly available software package \texttt{lenstronomy}\footnote{\url{https://github.com/lenstronomy/lenstronomy}} \citep{Birrer2015, lenstronomy, lenstronomy2}. We describe the assumptions for the mass, source light, and lens light distribution in Section~\ref{subsubsec:lens_model}. The reconstructed image predicted by the best-fitting lens model and the normalized residuals are displayed in Figure~\ref{fig:PSF_range_image_model}.

The fitting procedure involved iteratively optimizing the lens model parameters by maximizing the imaging likelihood with a Particle Swarm Optimization (PSO) algorithm, and reconstructing the PSF by proposing corrections to the PSF estimate from the previous step \citep{Chen2016, Shajib2019}. For our initial PSF estimate, we computed the mean of six PSF models generated with \texttt{stpsf}\footnote{\url{https://github.com/spacetelescope/stpsf}} \citep{webbpsf, stpsf} at wavelengths uniformly spaced in the same range as the one for the white-light image ($9320-9420\ \si{\angstrom}$). We show the initial PSF estimate and the model after iterative reconstruction in Figure~\ref{fig:PSF_range_image_model}. The reconstructed PSF is slightly more elongated in the direction of the IFS slices (i.e., along the $x$-axis), consistently with previous analyses  \citep[e.g.,][]{DEugenio024,Shajib2026}. To estimate the PSF uncertainties, we combined the variance map estimated by the reconstruction process \citep{Chen2016} and a term accounting for interpolation errors on the PSF grid, that we detail and motivate in Appendix~\ref{app:PSF_uncertainties}.

\subsubsection{Lens model}
\label{subsubsec:lens_model}

\begin{figure*}
    \centering
    \includegraphics[width=0.99\textwidth]{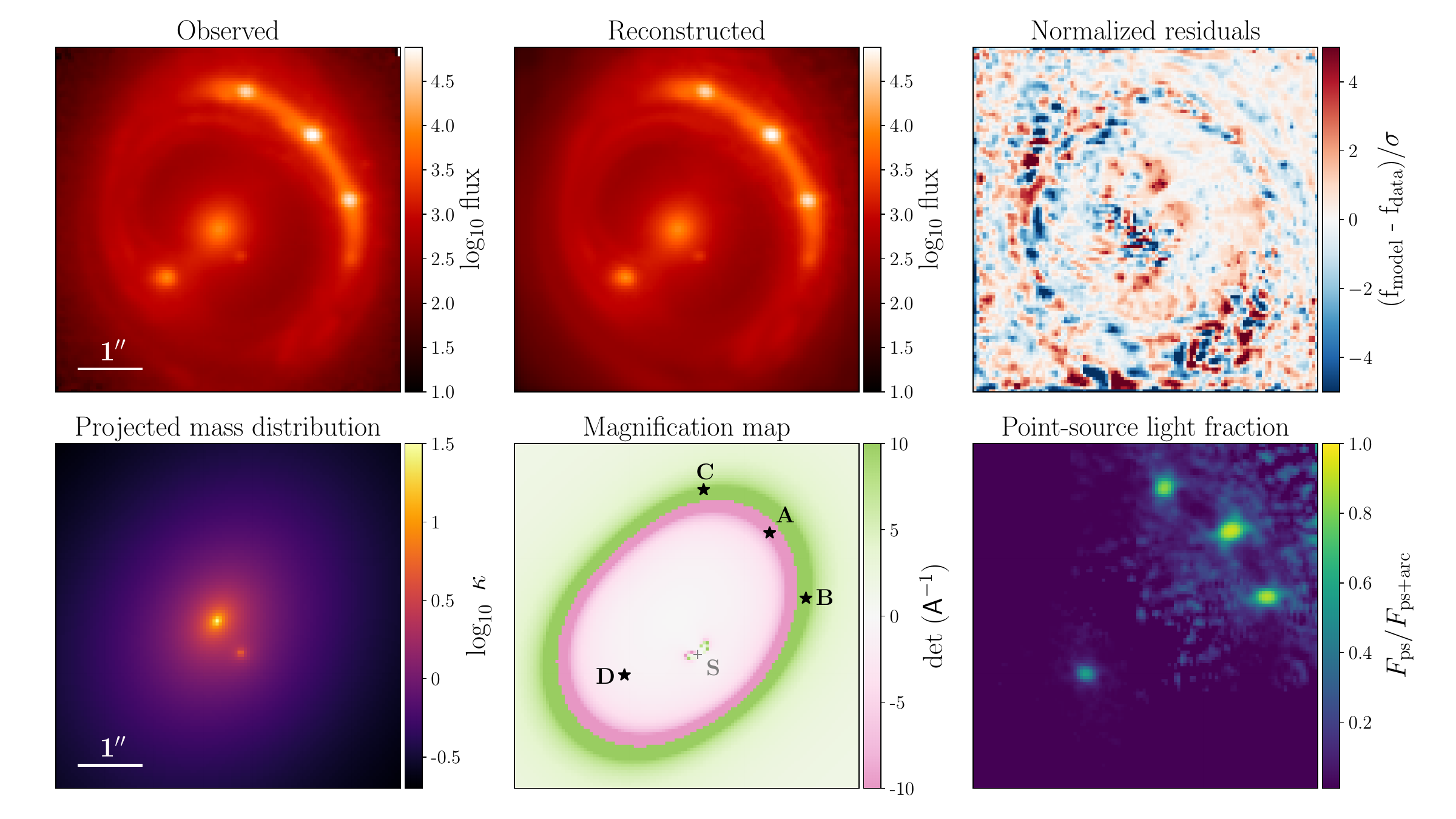}
    \caption{Image model for the detailed lens modeling step. \textbf{Top row,} from left to right: white-light image summed over the $8900-9650\ \si{\angstrom}$ wavelength range (in gray in Figure~\ref{fig:wavelength_ranges}), reconstructed light under the best-fit lens model, and normalized residuals. 
    \textbf{Bottom row,} from left to right: unitless convergence profile $\kappa$ (i.e., dimensionless projected surface mass density), corresponding model for the magnification $\mu = \det(\mathbf{A}^{-1})$ (where $\mathbf{A}$ is the lensing Jacobian), and fraction of unresolved quasar light compared to the total lensed light (extended emission + point-source flux).}
    \label{fig:spectra_range_image_model}
\end{figure*}

In the second step, keeping the PSF estimate fixed, we re-fit the lens model parameters on another white-light image, constructed in a larger wavelength range encompassing the narrow-line doublet, over which we wanted to extract and model the quasar image spectra. The range corresponding to the [\ion{S}{III}] doublet, shown in gray in Figure~\ref{fig:wavelength_ranges}, was chosen to be $8900-9650\ \si{\angstrom}$. The parameters were re-optimized using the PSO method again (without PSF reconstruction), then we estimated their posterior probability distribution using Markov Chain Monte-Carlo (MCMC) sampling. This second model fit allowed us to determine the lens properties “averaged” over the entire targeted wavelength range, while benefiting from the better PSF estimation from the restricted range (see Section~\ref{subsubsec:PSF_estimation}), implicitly neglecting the wavelength dependence of the PSF over the full range. We display the second white-light image in Figure~\ref{fig:spectra_range_image_model}, along with the best-fit lens model reconstruction, imaging residuals, convergence (projected surface mass distribution) and magnification prediction. In the following paragraphs, we list the assumptions that were used for the mass and light distributions, which are the same for this step and the previous one.

For the mass model, we assumed an elliptical power-law (EPL) surface mass distribution \citep{TessoreMetcalf2015} for the main deflector, and a singular isothermal sphere (SIS) profile for the nearby satellite galaxy, which is known to have an Einstein radius $\theta_E\sim 0\farcs2$ \citep{Suyu2013}. We also included a residual (constant) shear field \citep[also known as external shear, e.g.,][]{Kormann1994_shear, Keeton1997, Shajib2024} to represent deformations of the deflection field introduced by other masses along the line of sight.

The light distribution from the main deflector was modeled using two concentric elliptical Sérsic profiles \citep{Sersic1963, Sersic1968}, and the light from its satellite galaxy was included as a circular Sérsic with fixed half-light radius $R_{\rm e}=0\farcs01$ and Sérsic index $n=1$ \citep{Suyu2013, Shajib2026}. The source light distribution from the quasar host galaxy is known to be extremely complex for RXJ1131$-$1231 \citep[e.g.,][]{Claeskens2006, Suyu2013,Birrer2016, Shajib2026}. We find that both white-light images are well fit by using  a model with two non-concentric elliptical Sérsic profiles, plus two sets of shapelets \citep{Refregier2003, Birrer2015} with $n_{\rm max}=15$ and $n_{\rm max}=12$, with the centroid of each shapelet set joined with one of the Sérsics. This allows us to capture the complexity of the source at the location of the quasar, but also of a faint, secondary compact light source seen in the images (and in the residuals when only one profile is considered), offset from the point sources but still quadruply imaged (see Figure~\ref{fig:PSF_range_image_model}).

Finally, for the unresolved emission from the quasar, which we are aiming to isolate, we followed the standard method \citep[e.g.,][]{JWST_LQ_DM_III} and separately fit the quasar images as four point sources in the image plane, i.e., with positions and fluxes that are not directly determined by the lens model. This is critical for flux-ratio anomaly studies, since this ensures that the smooth lens model is not inferred using the observed image fluxes (which might be perturbed by substructure). The fitted image positions, however, are still used as constraints during the lens parameter sampling: by demanding that they map to a common location in the source plane (within a chosen uncertainty of $1$ mas), the lens equation is effectively enforced  \citep[e.g.,][]{Birrer2015, Gilman2020}. The point-source light fraction, i.e., the ratio between the point-source flux and the total lensed flux including the extended emission, directly relevant to our science case, is plotted in Figure~\ref{fig:spectra_range_image_model} for our best-fit lens model.

\subsubsection{Spectral extraction}
\label{subsubsec:spectra_extraction}

In the third step, we re-optimize, for each wavelength slice, the linear parameters of the lens model (this includes the amplitude parameters for each light component, e.g., the point source amplitudes and the coefficients of the Sérsic+shapelets parametrization for the extended emission), while fixing the PSF and the nonlinear lens model parameters (e.g., the mass model parameters). The resulting point source amplitudes as a function of wavelength can then be interpreted as the extracted spectra for the four images of the quasar's unresolved component.

To estimate the uncertainties on these spectra, we ran the extraction procedure $N_{\rm extr}\sim 100$ times, with nonlinear parameters randomly sampled from the MCMC posterior obtained during the lens modeling step (see Section~\ref{subsubsec:lens_model}). We used the best-fit PSF estimate from Section~\ref{subsubsec:PSF_estimation}, and added the estimated PSF variance map (see Appendix~\ref{app:PSF_uncertainties}) to the noise map around the quasar images, in order to also account for uncertainties in the PSF reconstruction. Finally, since the re-fitting of linear parameters amounts to a least-square optimization, we used the analytical estimator for their variance-covariance matrix \citep{Birrer2015} to randomly sample the error on the point source amplitudes, for each extraction. The resampled point source amplitudes were thus spread out according to uncertainties forwarded from each of the three steps of the modeling procedure. Since $N_{\rm extr}$ is relatively small (due to the extraction being computationally costly), we used robust estimators (the median and median absolute deviation) for the values and uncertainties of the final extracted spectra. These are displayed in Figure~\ref{fig:spectra_fit} for the range surrounding the [\ion{S}{III}] doublet.

\subsection{Quasar spectra model}
\label{subsec:spectral_model}

Given the source redshift \citep[$z_s=0.658$,][]{Sluse2003, Sluse2007}, there are two possible regions to target in the NIRSpec wavelength range (see Figure~\ref{fig:wavelength_ranges}): either $\sim 6200-6800\ \si{\angstrom}$ rest-frame to measure the [\ion{O}{I}], [\ion{N}{II}] and [\ion{S}{II}] narrow doublets \citep{VandenBerk2001}, or $\sim 9000-9600 \ \si{\angstrom}$ rest-frame to measure the narrow [\ion{S}{III}] doublet \citep{Riffel2006}. We attempted to measure the narrow-line flux ratios on the first range, but (i) [\ion{O}{I}] is too faint, even in aperture-integrated spectra (i.e., including the extended emission) (ii) [\ion{N}{II}] is completely blended with the broad and very strong H$\alpha$ emission, which is sensitive to microlensing (iii) after extraction of the quasar-only spectra (see Section~\ref{subsec:quasar_spectra_extraction}), the [\ion{S}{II}] lines are too faint, indicating that the emission seen in aperture-integrated spectra  is mostly due to extended emission rather than the quasar itself. We therefore measured the narrow-line flux ratios using only the prominent [\ion{S}{III}] doublet, and chose the $\sim 8900-9650 \ \si{\angstrom}$ wavelength range to extract and model the quasar image spectra.

\begin{figure*}
    \centering
    \includegraphics[width=0.997\textwidth]{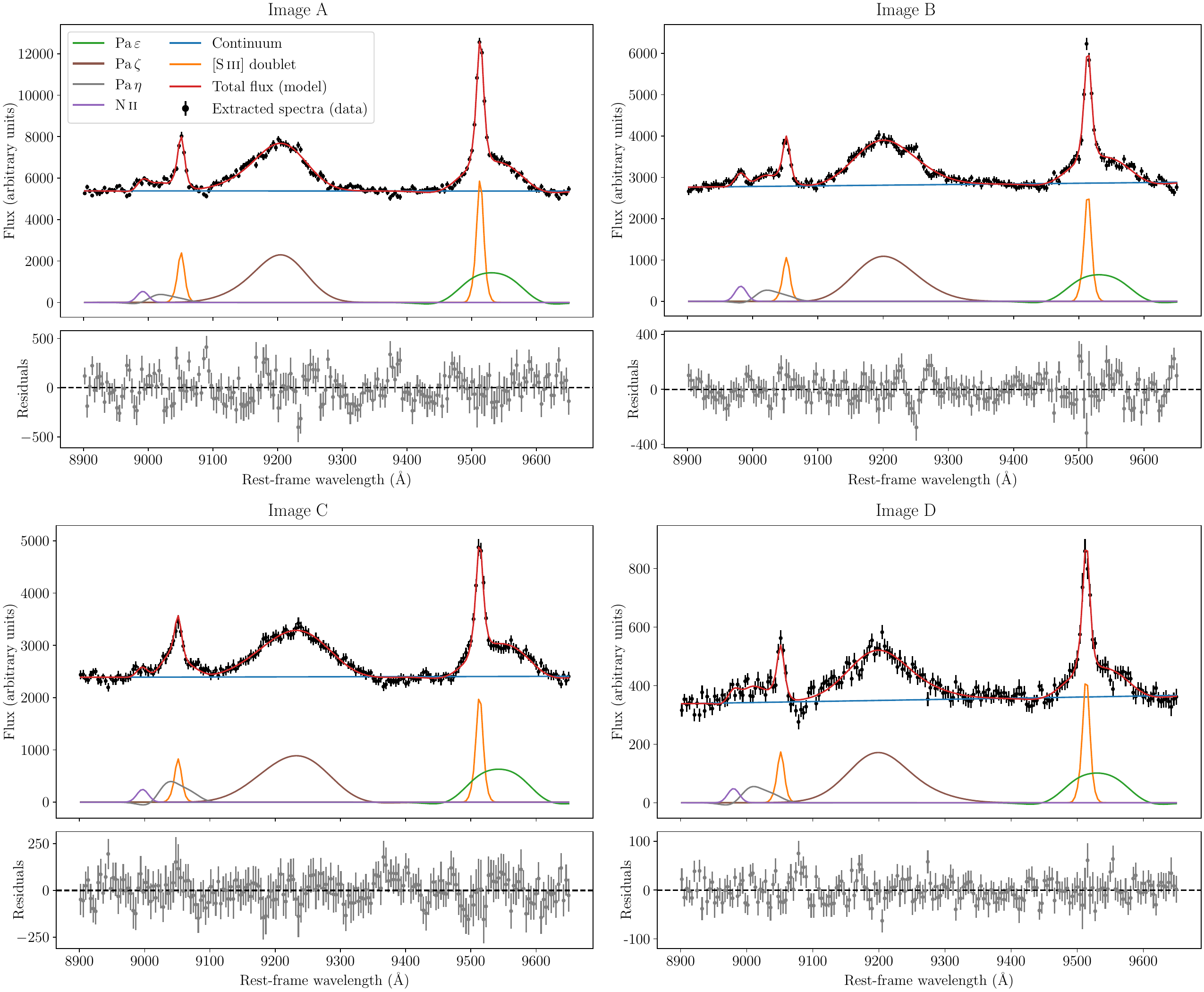}
    \caption{Best-fit joint model for the four extracted quasar spectra at images A, B, C, and D. The observed data points are shown in black, with error bars representing the estimated $\pm 1 \sigma$ uncertainties, forwarding the PSF and lens modeling uncertainties. Each spectral feature used in the model is displayed separately, with the total model flux in red. The absolute residuals are shown at the bottom of each panel.}
    \label{fig:spectra_fit}
\end{figure*}

The spectra for the four quasar images were jointly modeled in that wavelength range using the publicly available software package \texttt{lensqso-specfit}\footnote{\url{https://github.com/Lens-Dark-Matter/lensqso-specfit}}, that we developed for this work. This package handles the simultaneous modeling of multiple spectra by separating model parameters into nonlinear parameters (e.g., line shifts, widths, and shape parameters) which can be shared across all images (or fit separately to allow for image-specific microlensing deformations in some spectral components), and linear parameters (amplitudes/flux scalings) which are solved analytically for each image with linear least squares regression for efficient computation. The nonlinear parameters can be fit using the PSO implemented in \texttt{lenstronomy} \citep{lenstronomy, lenstronomy2}, constrained optimization by quadratic approximations \citep[\texttt{COBYQA},][]{COBYQA1, COBYQA2}, or exploration of the posterior with MCMC \citep[\texttt{emcee},][]{emcee}. We detail the implementation of this package in Appendix~\ref{app:lensqso_specfit}, presenting the formalism for the treatment of parameters in Appendix~\ref{app:lin_nonlin_params}, and the available spectral features and line profiles in Appendix~\ref{app:spec_feat}.

We modeled the spectra using the following components: 
\begin{itemize}
    \item a power-law continuum, 
    \item the narrow [\ion{S}{III}] doublet, with both lines described using the same Gauss-Hermite series with order $N=4$ \footnote{\label{c1c2zero} The first-order and second-order coefficients in the series are fixed to zero (see Appendix~\ref{app:line_profiles}).}, and with the relative amplitude of the two lines fixed at a theoretical line ratio $R(\lambda9533/ \lambda9071)=2.48$ \citep{Revalski2024},
    \item three prominent broad lines from the Paschen series that fall in the considered range (Pa\,$\varepsilon$, Pa\,$\zeta$, and Pa\,$\eta$), each represented independently using a Gauss-Hermite series with order $N=4$ \footref{c1c2zero},
    \item a faint line at $\sim 8990 \si{\angstrom}$ (we assumed it to be a known permitted \ion{N}{II} transition, based on the corresponding wavelength and the large transition probability), most prominently seen in image B but improving the fits for all images, modeled as a simple Gaussian.
\end{itemize}
The profiles used in the final model for each of these spectral features are summarized in Table~\ref{tab:spec_feat}, along with the central vacuum rest-frame wavelength for the emission lines. In the model, the lines are allowed to be redshifted with respect to the mean source redshift ($z_s=0.658$) and relative to each other, to account for line-of-sight velocities.

\begin{table}
\caption{\label{tab:spec_feat} Summary of the spectral features used for the best-fit model of the quasar spectra in the $8900-9650\ \si{\angstrom}$ range. The values for the rest-frame vacuum wavelength $\lambda_{\rm vac}$ of each emission line are taken from the NIST database \citep{NIST}. For the conventions used for the Gauss-Hermite series, see Appendix~\ref{app:line_profiles}.}
\centering
\begin{tabular}{lcr}
\hline \vspace{-0.3cm}
\\
\textrm{Spectral feature}&
\textrm{$\lambda_{\rm vac}$ ($\si{\angstrom}$)}&
\textrm{Model profile}\\
  \hline  \hline 
Continuum & - & Power-law \\
\hline
[\ion{S}{III}] doublet & 9 071.1   & Gauss-Hermite ($N=4$) \\
 &  9 533.2 &  \\
\hline
Paschen-$\varepsilon$ (Pa\,$\varepsilon$) $^*$ & 9 548.8 & Gauss-Hermite ($N=4$) \\
Paschen-$\zeta$ (Pa\,$\zeta$) $^*$ & 9 232.2 & Gauss-Hermite ($N=4$) \\
Paschen-$\eta$ (Pa\,$\eta$) $^*$ & 9 017.8 & Gauss-Hermite ($N=4$) \\
\ion{N}{II} & 8 988.6 & Gaussian \\
  \hline  \hline
\end{tabular}\\
\footnotesize{$^*$ In the literature, the Paschen series lines are sometimes named based on the upper energy level (Pa\,8, Pa\,9, Pa\,10, etc.~instead of Pa\,$\varepsilon$, Pa\,$\zeta$, Pa\,$\eta$, etc.)}
\vspace{-0.1cm}
\end{table}

For the joint modeling, we need to account for the fact that the images are expected to be magnified relative to each other due to gravitational lensing. Continuum and broad emission features, in addition, are intrinsically variable on small timescales, and are susceptible to differential magnification by stars in the plane of the lens galaxy \citep[e.g,][]{Keeton2006, Sluse2007, Anguita2008}. This can affect the shape of these emission features, since the bluer parts of the continuum and the broad-line emission at higher velocity are emitted from systematically smaller regions, which are more sensitive to microlensing \citep[e.g.,][]{Abajas2002, Blackburne2011, Sluse2012, Bate2018, Fian2018, Fian2024, Hutsemekers2024}. To account for these effects, the slope and amplitude of the continuum was inferred separately for each image. Similarly, we initially fit the broad line shape parameters (width, relative Gauss-Hermite coefficients $c_i/c_0$) independently for each image, but we found that, for Pa\,$\varepsilon$ and Pa\,$\eta$, it did not significantly improve the fit or change the final flux-ratio values compared to a model where these parameters are shared between the four images. Thus, in the final model, only Pa\,$\zeta$ had shape parameters that were allowed to vary between images. Finally, since the narrow-line emission is not sensitive to microlensing, we assumed shared shape parameters between the four images, with only the flux scalings fit separately. 

We show the best-fitting joint model for the four images in Figure~\ref{fig:spectra_fit}, along with the decomposition into the different spectral components. The residuals (also displayed in Figure~\ref{fig:spectra_fit}) are, after normalization by the estimated uncertainties, consistent with a standard normal distribution (sample mean $\approx 0$ and sample variance $\sim 1$).

We note that we attempted to model the extracted spectra with many different variations around the assumptions presented here, that did not improve the fit or alter the narrow-line flux-ratio measurements in a significant way. We report the [\ion{S}{III}] flux ratios relative to image A for a selection of those alternative spectral models in Appendix~\ref{app:FRs_alt_specmodels}. In particular, we tried using other line profiles (Voigt and Gauss-Hermite with different orders—see Appendix~\ref{app:line_profiles}), changing the (fixed) value for the relative amplitude between the two lines in the [\ion{S}{III}] doublet, and freeing some of the parameters to fit them independently between images. We also implemented a semi-empirical template for \ion{Fe}{II} and \ion{Mg}{II} near-infrared (NIR) emission lines, taken from \cite{Garcia-Rissmann2012} and described in Appendix~\ref{app:spec_feat}. The most prominent feature in the fitted range would be a \ion{Fe}{II} bump at $\sim9200 \ \si{\angstrom}$ ; while the observed spectral feature at this wavelength could be a blend of Pa\,$\zeta$ and of the \ion{Fe}{II} bump, we are only truly interested in the wavelength regions around the [\ion{S}{III}] lines, which should receive a negligible contribution from \ion{Fe}{II} lines according to the template. Therefore, since a single line is able to fit the spectra reasonably well around $\sim9200 \ \si{\angstrom}$, we did not include the template lines in our final model. Because the measured narrow-line flux ratios did not change significantly during these various tests, we conclude that the values presented in Section~\ref{sec:results} are robust to spectral modeling assumptions.

\section{Results \& comparison with other measurements.}
\label{sec:results}

In this section, we present and comment our final measurements for the narrow-line flux ratios of RXJ1131$-$1231, putting them in perspective with other relevant quantities. In Section~\ref{subsec:FR_values}, we start by discussing all the results that can be inferred from the JWST/NIRSpec data alone, comparing the narrow-line, broad-line, continuum, and lens-model-predicted flux ratios. In Section~\ref{subsec:literature_comparison}, we leverage the fact that RXJ1131$-$1231 has been extensively studied in the literature,  and confront our new measurements to previous flux-ratio values obtained across a wide range of emission regions and wavelengths.

\vfill

\subsection{Final values \& comparison with lens model predictions}
\label{subsec:FR_values}

\begin{table*}
\caption{\label{tab:results_FR} Measured flux ratios and astrometric offsets between the point-like quasar images. By convention, we compare Image 1 relative to Image 2 (i.e., we report $\vec{x}_1 - \vec{x}_2$ and $f_1/f_2$). The flux-ratio uncertainties are estimated by MCMC exploration of the posterior for the joint spectral model. We separate the values between the 3 cusp images (A, B, and C), which are particularly relevant for dark substructure inference, and those involving the counterimage (D).}
\centering
\begin{tabular}{cccccccc}
\hline
\textrm{Image 1} &  \textrm{Image 2} &
\textrm{$\Delta\rm RA$ ($^{\prime\prime}$) $^*$ }&
\textrm{$\Delta\rm DEC$ ($^{\prime\prime}$) $^*$ } & \textrm{Lens model prediction $^\dagger$} & [\ion{S}{III}] flux ratio  & Continuum flux ratio & Pa\,$\varepsilon$ flux ratio  \\
  \hline  \hline 
B & A & $\phantom{-}0.007$ & $\phantom{-}1.194$ & $0.64 \pm 0.01$ & $0.45 \pm 0.02$ & $0.53 \pm 0.02$ & $0.45 \pm 0.03$ \\
C & A  & $-0.592$&  $-1.111$ & $0.54 \pm 0.01$ & $0.35 \pm 0.02$ & $0.45 \pm 0.02$ & $0.44 \pm 0.02$ \\
C & B & $-0.599$ & $-2.305$ & $0.84 \pm 0.01$ &  $0.78 \pm 0.04$ & $0.85 \pm 0.04$ & $0.99 \pm 0.06$ \\
\hline
D & A & $-3.120$ & $\phantom{-}0.873$ &  $0.061 \pm 0.001$ & $0.070 \pm 0.004$ & $0.066 \pm 0.004$ & $0.071 \pm 0.006$ \\
D & B  & $-3.127$ & $-0.321$ & $0.094 \pm 0.001$ & $0.16 \pm 0.01$ & $0.13 \pm 0.01$ & $\phantom{0}0.16 \pm 0.015$  \\
D & C  & $-2.529$ &  $\phantom{-}1.984$ & $0.112 \pm 0.001$ & $\phantom{0}0.20 \pm 0.015$ & $0.15 \pm 0.01$ & $\phantom{0}0.16 \pm 0.015$  \\
  \hline  \hline
\end{tabular}\\
\footnotesize{$^*$ We adopt typical astrometric uncertainties of $\sim 0\farcs01$ based on the MCMC posterior from the lens modeling step. Uncertainties due to PSF reconstruction should be comparable, of the order of a fraction of a pixel \citep{Chen2021}.\\
$^\dagger$ For the EPL+shear+SIS lens model presented in Section~\ref{subsubsec:lens_model}.}
\vspace{-0.1cm}
\end{table*}

\begin{figure*}
    \centering
    \includegraphics[width=\textwidth]{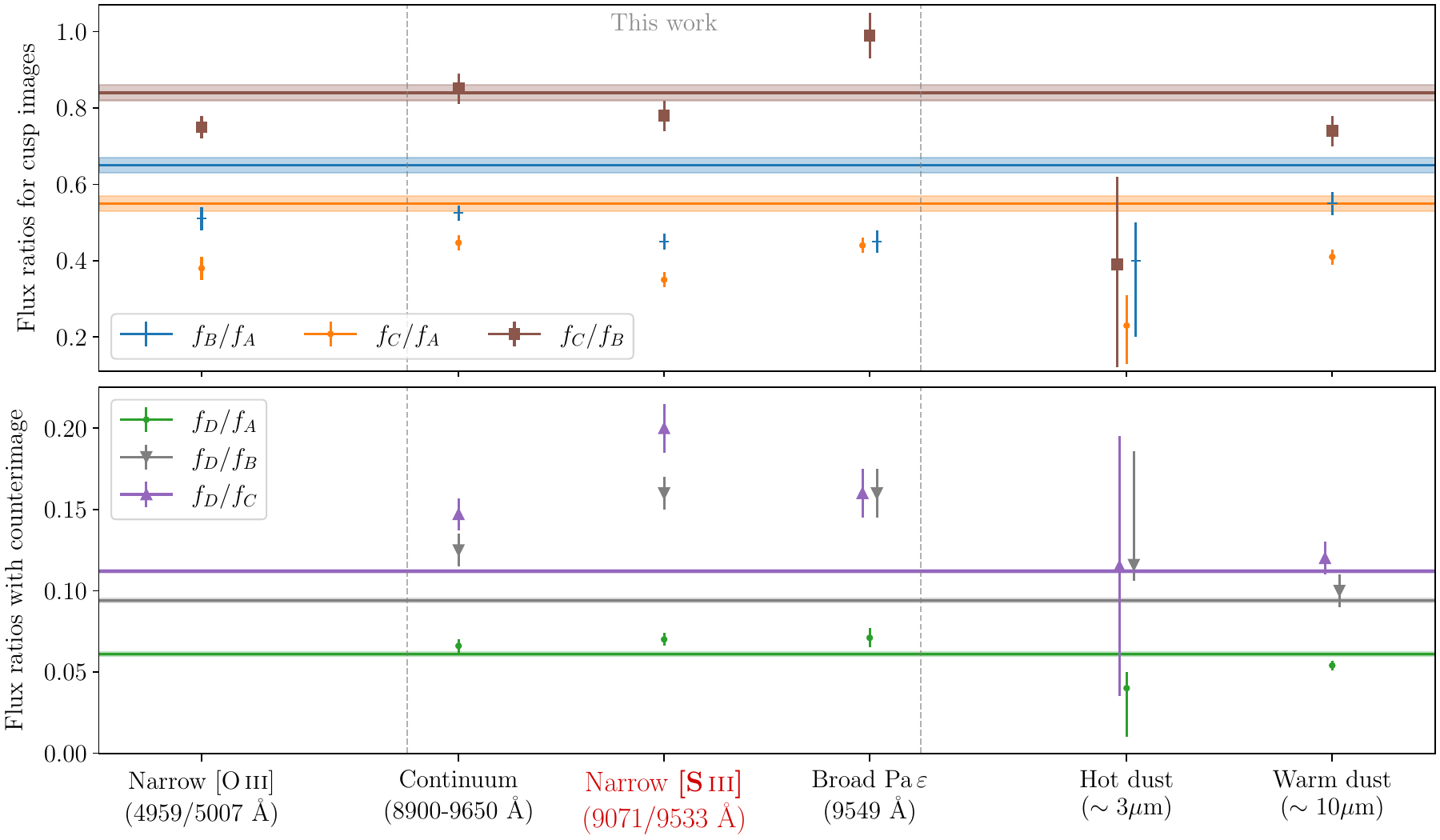}
    \caption{ Comparison between the flux ratios measured in this work (for the narrow [\ion{S}{III}] doublet, the broad Pa\,$\varepsilon$ line, and the continuum over the $8900-9650 \si{\angstrom}$ range), the magnification ratios predicted by the EPL+shear+SIS lens model described in Section~\ref{subsubsec:lens_model} (displayed as filled bands representing 95\% credible intervals), and some flux-ratio values found in the literature. The narrow [\ion{O}{III}] doublet measurements are taken from \protect\cite{Sluse2007}, and the hot/warm dust measurements from \protect\cite{JWST_LQ_DM_III}.}
    \label{fig:FR_comparison}
\end{figure*}

The narrow-line flux ratios measured from the [\ion{S}{III}] doublet are reported in Table~\ref{tab:results_FR}, along with the astrometric offsets between images determined during the lens modeling step. The flux ratios for all pairs of images are also displayed in Figure~\ref{fig:FR_comparison}. Some of the flux ratios are correlated, in particular when the same image flux is used to calculate the two considered ratios; to allow for an easier and more legible comparison between features and with ratios reported in the literature, however, we focus on individual values in this section and discuss these correlations in Appendix~\ref{app:correlation_FRs} (in particular, we present a full joint posterior for the [\ion{S}{III}] flux ratios). The measurement uncertainties on individual ratios are comparable in magnitude ($\sim 5 \%$) to those achieved with warm dust flux ratios from JWST/MIRI \citep{JWST_LQ_DM_III} in the presence of a bright quasar host, and to the expected accuracy obtained when modeling simulated systems with realistic extended narrow-line emission \citep{PerezMendoza2026}. This demonstrates that JWST/NIRSpec IFS narrow-line measurements can reach competitive precision for substructure inference. 

The measured [\ion{S}{III}] flux ratios reveal a clear anomaly relative to the magnification ratios predicted at the quasar image positions by the smooth lens model from Section~\ref{subsubsec:lens_model}, which are reported in Table~\ref{tab:results_FR} and displayed as 95\% credible intervals in Figure~\ref{fig:FR_comparison} (for a full joint posterior, see Appendix~\ref{app:correlation_FRs}). 
We emphasize that the values and uncertainties on the predicted magnification ratios are conditional on the adopted EPL+shear+SIS smooth lens model. We find that, for this lens model, the ratios $f_B/f_A$ and $f_C/f_A$ both deviate significantly from predictions ($\sim 9\sigma$ for the individual ratios), while $f_C/f_B$ is consistent with expectations. Flux ratios involving the counterimage (D) also have discrepancies with the predicted magnification ratios (in particular, $f_D/f_B$ and $f_D/f_C$). Typically, however, the flux ratios between the three merging images of the cusp (A, B, and C) are examined more closely since they probe the lens potential at comparable projected radii from the deflector center. As a consequence, they are less subject to systematic uncertainties from the lens model (e.g., they depend less on an accurate determination of the radial slope) and are therefore more robust tracers of the substructure perturbations. The anomalous nature of $f_B/f_A$ and $f_C/f_A$ suggest that image A is more magnified than predicted in our lens model, either due to the presence of DM substructure or to complexity in the mass distribution not captured by the EPL+shear+SIS model \citep[e.g.,][]{EvansWitt2003, Congdon2005, Xu2015, Powell2022, PaugnatGilman2025}. These two scenarios can only be discriminated at a statistical level \citep[e.g.,][]{Gilman2024}; determining the most likely configuration would require a full DM substructure inference \citep[like in][]{JWST_LQ_DM_II, JWST_LQ_DM_III, JWST_LQ_DM_IV}, which is beyond the scope of this work.

For completeness, we also computed flux ratios measured from a broad Paschen emission line and from the continuum. Among the detected broad lines, we used only Pa\,$\varepsilon$, since the \ion{N}{II} and Pa\,$\eta$ lines are too faint (and partially blended), and Pa\,$\zeta$ is possibly blended with \ion{Fe}{II}/\ion{Mg}{II} features and has different shape parameters between images (see Section~\ref{subsec:spectral_model}). Continuum flux ratios were measured by integrating the best-fitting continuum model over the same wavelength range used for the spectral fitting. As shown in Figure~\ref{fig:FR_comparison}, some of the broad-line and continuum flux ratios differ substantially from the narrow-line values. This is expected given that the broad-line region and accretion disk are sensitive to significant differential microlensing \citep[e.g,][]{MoustakasMetcalf2003}.

We expect that our final measurement uncertainties ($\sim5\%$) are large enough to absorb any remaining systematics.
While lens modeling assumptions directly impact the predicted magnification ratios, they are considerably less important for the measured flux ratios, since they only weigh in through the separation of unresolved nuclear emission from extended components, and the nuclear contribution dominates near the core of the PSF, in particular for the cusp images (see bottom right panel of Figure~\ref{fig:spectra_range_image_model}). Even in our case, where extended emission is prominent comparatively to other quads, tests on simulated data demonstrate that narrow-line flux ratios can be recovered to better than 5\% accuracy \citep{PerezMendoza2026}. Uncertainties associated with the wiggle correction (Appendix~\ref{app:wiggle_correction}) are explicitly propagated through our error budget, and remaining systematics in the spectral baseline should average out when integrating the fluxes over the full PSF. Uncertainties due to the PSF reconstruction and interpolation (Appendix~\ref{app:PSF_uncertainties}) are accounted for, and residual effects (e.g., PSF chromaticity in the relatively small wavelength range) are expected to be negligible: common perturbations to the PSF shape shift all fluxes coherently and thus largely cancel out in the ratios.

\subsection{Comparison with the literature}
\label{subsec:literature_comparison}

Prior flux-ratio measurements for RXJ1131$-$1231 span a range of emission regions and wavelengths, and differ in their degree of relevance for direct comparison with our results. First, the flux ratios for this lensed system are known to be highly variable in broad-band optical monitoring \citep[R-band photometry,][]{Tewes2013, Millon2020}, mm-wavelength monitoring with ALMA \citep{Rybak2017, Rybak2025}, or X-ray monitoring with Chandra \citep{Chartas2012}. These measurements are indeed dominated by emission from the compact accretion disk and corona, which are strongly affected by microlensing, and further complicated by the time delays between images, which are comparable to the variability timescale for these emission regions. They are therefore not directly comparable to narrow-line flux ratios, which are immune to both effects.

Narrow-line measurements are reported by \cite{Sugai2007}, who used the IFS mode of the Kyoto 3DII spectrograph on the Subaru telescope to observe images A, B, and C simultaneously over a wavelength range encompassing the [\ion{O}{III}] $4959/5007 \si{\angstrom}$ narrow-line doublet. They present [\ion{O}{III}] flux-ratio values ($f_A/f_B = 1.63^{+0.04}_{-0.02}$ and $f_C/f_B = 1.19^{+0.03}_{-0.12}$) that are consistent with smooth lens model predictions (i.e., no flux-ratio anomaly), but their calculation relies on spectra integrated over $0\farcs77$-diameter apertures. Since these would receive a significant contribution from the extended emission (which corresponds to much larger physical sizes and therefore is not sensitive to millilensing), it is likely that their flux-ratio values are not anomalous due to the extended emission diluting perturbations from substructure that are observed when considering only the unresolved emission. To confirm this, we re-iterated our spectral modeling procedure on aperture-integrated spectra (using a radius of $8$ pixels, i.e., $0\farcs8$-diameter apertures), and found values for the [\ion{S}{III}] flux ratios ($f_A/f_B = 1.54 \pm 0.08$ and $f_C/f_B = 1.17 \pm 0.05$) consistent with \cite{Sugai2007}.
This highlights the importance of the lens modeling step to isolate the unresolved emission when extracting the spectra (see Sections~\ref{subsubsec:lens_model} and \ref{subsubsec:spectra_extraction}).

A more direct comparison is possible with \cite{Sluse2007}, who used long-slit optical and NIR spectroscopic data from the Very Large Telescope, modeled using a multi-component spectral decomposition, to extract flux ratios between the cusp images for the continuum, broad lines (H$\alpha$, H$\beta$), and narrow lines, in particular the [\ion{O}{III}] doublet. They measure $f_A/f_B = 1.97 \pm 0.03$ and $f_C/f_B = 1.33 \pm 0.02$ for the [O III] ratios, which are roughly consistent with our measurements. They find substantial discrepancies with continuum and broad-line ratios, that they attribute to microlensing de-amplification of images A and C relative to image B.  By directly matching the spectra with a separation between microlensed and non-microlensed contributions, they also determine microlensing-free amplification ratios $f_A/f_B = 2.1$ and $f_C/f_B = 1.4$, which match our values as well. While they also do not correct explicitly for extended emission, they use a point-like source and a PSF model to extract their quasar spectra, instead of integrating over apertures. This mitigates the contribution from the resolved component since unresolved light dominates near the core of the PSF (see Figure~\ref{fig:spectra_range_image_model}), and likely explains why their values are comparable to ours. With this caveat in mind, we report their [\ion{O}{III}] flux-ratio values in Figure~\ref{fig:FR_comparison}.

Finally, it is particularly interesting to consider the mid-infrared flux ratios measured by \cite{JWST_LQ_DM_III} using JWST/MIRI multiband photometry, as part of the JWST lensed quasar dark matter survey \citep[see also][]{JWST_LQ_DM_I, JWST_LQ_DM_II, JWST_LQ_DM_IV}. Their method relies on a simple model of the quasar spectral energy distribution (SED) to separate contributions from (i) continuum emission, 
(ii) a “hot” \citep[$\sim1200$ K,][]{Hernan-Caballero2016} dust component peaking at rest-frame $\sim3 \mu$m, associated with the dust sublimation zone—which can be quite compact ($0.05-0.2$~pc) and thus sensitive to microlensing, and
(iii) a “warm” \citep[$\sim400$ K,][]{Hernan-Caballero2016} dust component dominating the SED at rest-frame $\sim 10 \mu$m, emitted from a larger region ($0.5-10$~pc) and thus not susceptible of being microlensed. We display their final values for the hot and warm dust flux ratios in Figure~\ref{fig:FR_comparison}. 

\begin{figure*}
    \centering
    \includegraphics[width=1.01\textwidth]{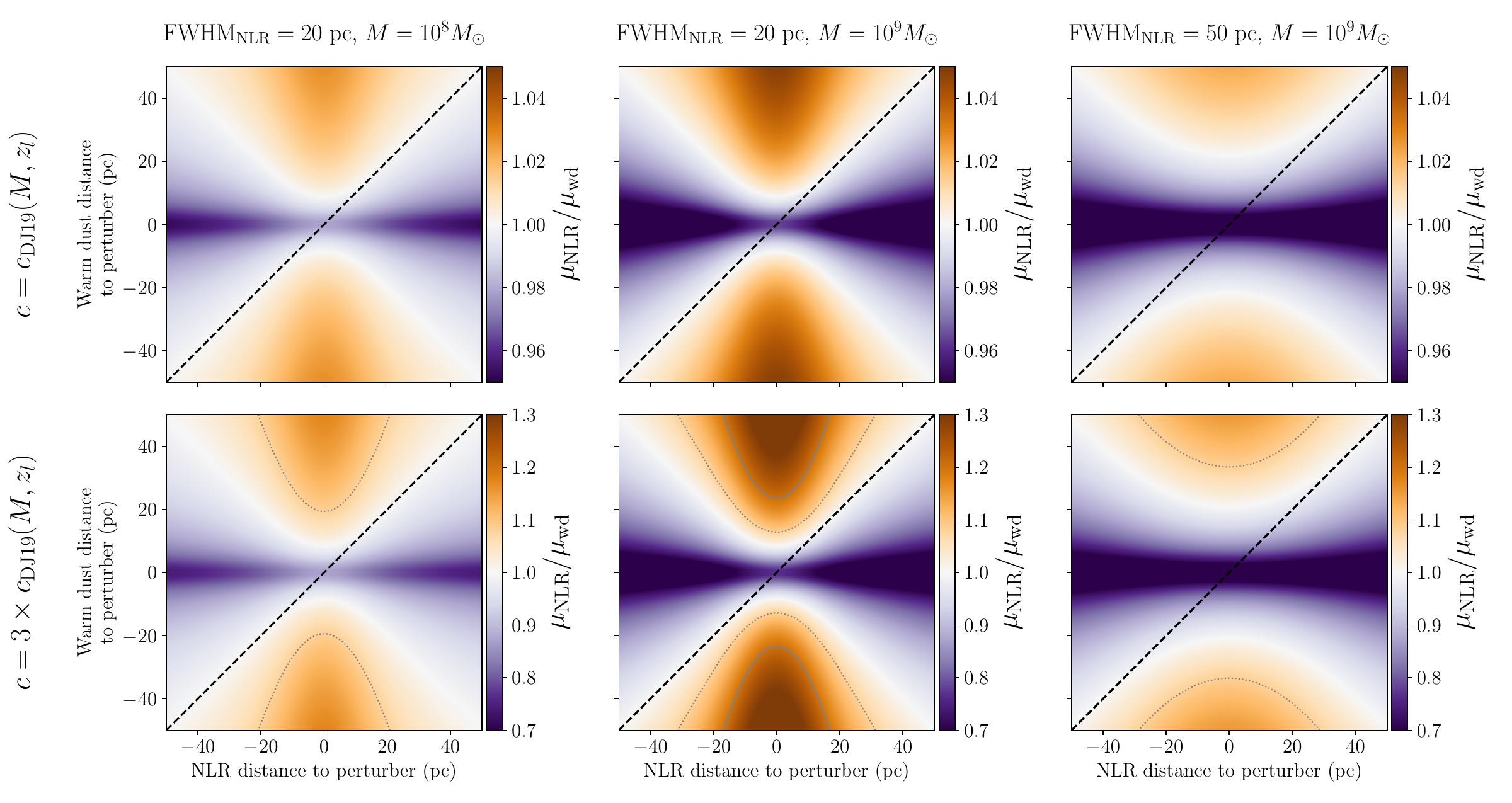}
    \caption{Relative magnification $\mu_{\rm NLR}/ \mu_{\rm wd}$ between the NLR and the warm dust, as a function of the offset of both emission regions relative to the center of a perturbing NFW halo. The areas in orange (respectively, in purple), where the ratio $\mu_{\rm NLR}/ \mu_{\rm wd}$ is $>1$  (resp., $<1$) correspond to parameter space where the narrow-line emission (resp., the warm dust emission) is preferentially magnified by the perturber. We assume Gaussian light distribution for the profiles of both regions, with a FHWM of 5~pc for the warm dust \protect\citep{JWST_LQ_DM_III}. In order to illustrate the impact of the other parameters, we vary the FHWM of the NLR \protect\citep[20~pc or 50~pc,][]{Muller-Sanchez2011}, the perturbing halo mass ($M=10^8 M_\odot$ or $M=10^9 M_\odot$) on each column, and the perturbing halo concentration ($c/c_{\rm DJ19} =1$ or 3, where $c_{\rm DJ19}(M, z_l)$ is the concentration-mass relation from \protect\cite{DiemerJoyce2019}) between the top and bottom rows (note the change in scale for the colormap). The dashed black line indicates concentric emission regions (i.e., the two are equally offset from the NFW halo). The dotted and solid gray lines are isocontours at $\mu_{\rm NLR}/ \mu_{\rm wd}=1.1$ and 1.2, respectively.}
    \label{fig:mag_NLR_wd_with_offset}
    \vspace{0.1cm}
\end{figure*}

The general behavior of the warm dust flux ratios is comparable to our narrow-line measurements ($f_B/f_A$ and $f_C/f_A$ are smaller than the lens model predictions, $f_C/f_B$ is consistent), which confirms the presence of the flux-ratio anomaly. There are, however, marginal discrepancies ($2-3~ \sigma$ differences on individual ratios, see Appendix~\ref{app:correlation_FRs} for a comparison of the joint posteriors) between the [\ion{S}{III}] and warm dust values in the cusp. Having carefully propagated all identified sources of uncertainty, we consider it worthwhile to explore physical explanations beyond simply attributing these differences to systematic errors.

Since both emission regions are different in size, we actually do not expect identical amplitudes for flux-ratio perturbations from substructure. In particular, the more compact warm dust region is sensitive to individual halos $\gtrsim 10^7 M_\odot$ and to the collective effect of substructure down to $\sim 10^6 M_\odot$ \citep{JWST_LQ_DM_I, JWST_LQ_DM_II}, while the more extended NLR predominantly probes more massive halos. For instance, if the two regions are modeled as concentric Gaussians with respective FWHM $\sim 5 $~pc and $\sim 50-100$~pc, the warm dust flux ratios should be more anomalous than the narrow-line values (see Figure~1 from \cite{JWST_LQ_DM_I} for an example with a $\sim 10^7M_\odot$ halo). Here, however, $f_B/f_A$ and $f_C/f_A$ are slightly more anomalous in the narrow-line ratios, which hints at a more complex geometry.

We tested the plausibility of one physical explanation: a spatial offset between the two emission regions \citep[already considered in][]{JWST_LQ_DM_I}, since the distance between the center of the nuclear narrow-line region and the quasar accretion disk is sometimes observed to be of the order of tens of parsecs in low-redshift quasars \citep{Muller-Sanchez2011}, i.e., comparable to the FWHM of the NLR. We employed a very simple toy model, assuming Gaussian emission regions lensed by a field DM halo, placed at the same redshift as the main lensing galaxy ($z_l=0.295$), and with a 3D mass distribution following a Navarro-Frenk-White (NFW) profile \citep{NFW1997}. We then calculated the magnification of the total NLR flux relative to the total warm dust flux ($\mu_{\rm NLR}/\mu_{\rm wd}$), as a function of the offset of each region relative to the center of the perturbing halo. We adopt a FWHM of $5$~pc for the warm dust and plot the results in Figure~\ref{fig:mag_NLR_wd_with_offset}, varying the FWHM of the NLR ($20$~pc or $50$~pc), the mass of the perturbing halo ($M=10^8 M_\odot$ or $M=10^9 M_\odot$), and its concentration ($c/c_{\rm DJ19} =1$ or 3) relative to the CDM concentration-mass relation $c_{\rm DJ19} (M, z_l)$ from \cite{DiemerJoyce2019}. If we assume that the flux-ratio anomaly is caused by a DM substructure perturbing image A, the observed difference between the narrow-line and warm dust measurements in the cusp can be explained by such a physical scenario if $\mu_{\rm NLR}/\mu_{\rm wd} \approx 1.2$ (or $\mu_{\rm NLR}/\mu_{\rm wd} \approx 1.1$ to bring the values within $1\sigma$ of one another). This is more easily achieved when the NLR is more compact, when the halo is more massive, and when it is more concentrated, since those factors increase the additional magnification from substructure.

We remark that there is an upper limit to the offset between the warm dust and the NLR (i.e., the distance to the dashed black line in Figure~\ref{fig:mag_NLR_wd_with_offset}), that can be set using the observed astrometry of the point sources. First, when we compare the relative astrometry measured in our lens model (see Table~\ref{tab:results_FR}) to the values reported by \cite{JWST_LQ_DM_III} for the warm dust, we find differences $\sim10-20$ mas, i.e., within the estimated uncertainties. We also performed the following test to see if offsets were detectable in our data: keeping the PSF estimate and lens mass model fixed (to the ones obtained in Section~\ref{subsec:quasar_spectra_extraction}), we re-optimized the light model parameters (and in particular, the point-source positions) on two white-light images where the quasar spectrum is dominated by broad-line emission ($9100-9300 \si{\angstrom}$) and by narrow-line emission ($\pm10\si{\angstrom}$ around the peaks of the two lines in the [\ion{S}{III}] doublet), respectively. We find that the point-source positions are consistent within the $\sim10$~mas uncertainties in the image plane, which maps back to $\sim25$ pc in the source plane according to the lens model. Therefore, physical offsets between emission regions in RXJ1131$-$1231 should not significantly exceed $\sim 50$~pc. As shown in Figure~\ref{fig:mag_NLR_wd_with_offset}, even under such constraints, the relative magnification between the NLR and the warm dust can be relevant depending on the properties of the perturbing halo and the size of the regions.

We emphasize that (i) this is only one possible explanation for the observed marginal differences, and (ii) we presented a very simplified toy model, only intended to discuss the impact of spatial offsets qualitatively, to which we append some important caveats. First, we only considered a single field halo, while in reality, we expect many field halos and main deflector subhalos along the line of sight, potentially perturbing several images simultaneously \citep[e.g.,][]{Gilman2019, Gilman2024}. Furthermore, the lens and source in RXJ1131$-$1231 are located at low redshifts \citep[relative to the lensed quasar population,][]{OguriMarshall2010}, so the line-of-sight volume is relatively small and the subhalos are more likely to contribute to DM substructure perturbations. Interpretation is therefore more complex since subhalos evolve within their host halo under multiple processes \citep[e.g., tidal stripping, tidal heating, and dynamical friction,][]{Du2024, Du2025} and have different properties than field halos. For instance, in our toy model, the “boosted” concentration  ($c = 3 \times c_{\rm DJ19}$) can be seen as an upper limit for CDM field halos (it corresponds to $+3\sigma$ for the \cite{DiemerJoyce2019} relation, which has a scatter of 0.16 dex), but subhalos that appear near the Einstein radius and have retained most of their mass since infall can be significantly more concentrated than average \citep{Du2025}. Finally, we note that spatial offsets can also result in small difference in flux ratios if the anomaly is caused by complexity in the mass distribution rather than DM substructure: since the two emission regions would be positioned slightly differently relative to the caustic, they would not be magnified by the exact same amount. Extensively exploring the space of possible configurations to determine the most likely scenario is a complex endeavor that we leave to future work. 

\vspace{-0.4cm}
\section{Summary \& future prospects}
\label{sec:conclusion}

We have presented the first JWST/NIRSpec IFU measurements of flux ratios for a multiply imaged quasar, which are also the first measurements of this nature relying on the [\ion{S}{III}] narrow-line doublet. These values are particularly interesting for studies employing flux-ratio anomaly statistics as a means to constrain the properties of DM substructure. We have described a procedure to extract the spectra for the unresolved emission in each quasar image, removing the contribution from extended emission, which can dilute perturbations from substructure. We also have introduced the software package \texttt{lensqso-specfit}, designed to jointly model the multiple quasar spectra with the same features in a computationally efficient way, with the option to share parameters between images. We find that the targeted system, RXJ1131$-$1231, possesses significant discrepancies between the measured narrow-line flux ratios and the predictions from a standard (EPL+shear+SIS) smooth lens model, in particular between the three images in the cusp. Our results are consistent with previous narrow-line measurements reported in the literature, and agree within 2-3 $\sigma$ with warm dust values measured with JWST/MIRI. The small differences hint at the narrow-line fluxes being more anomalous than the mid-IR fluxes. While this could simply be a product of systematic uncertainties, the observed differences are also consistent with one plausible physical scenario: if the NLR and warm dust emission region are slightly offset from one another, perturbing DM substructure can preferentially magnify the NLR, especially if the halo is massive, concentrated, and if the bulk of narrow-line emission is relatively compact. We have illustrated this with a very simple toy model, and plan to explore this more rigorously in future work by performing a full substructure inference using time-tested forward-modeling techniques \citep[e.g.,][ and references therein]{JWST_LQ_DM_IV}.

More generally, this work illustrates the interest of using dual sources for flux-ratio anomaly statistics, that is, leveraging both narrow-line and warm dust flux-ratio measurements to constrain the properties of DM substructure in lensed quasars. The benefit is that, while both emission regions are not subject to microlensing, they have different physical sizes and are therefore sensitive to millilensing by DM halos on different mass scales: the warm dust is more compact and probes halo masses down to $\sim 10^6 M_\odot$ \citep[e.g.,][]{JWST_LQ_DM_II}, while the narrow-line region is mostly sensitive to the most massive halos. 
Dual-source flux ratios can therefore help break degeneracies between some substructure configurations. Several lensed quasars already have narrow-line flux ratios that were measured using the HST WFC3/IR grism \citep{Nierenberg2017, Nierenberg2020}, with some that overlap with the JWST/MIRI warm dust sample \citep{JWST_LQ_DM_III}. We plan to increase the number of systems with both measurements in the future.

Our method is indeed generalizable to other lensed quasar systems with existing or upcoming IFS data, and to any sufficiently bright narrow-line doublet falling within the observed spectral range. The [\ion{S}{III}] $9071/9533\ \si{\angstrom}$ doublet targeted here complements other narrow-line features (in particular, the commonly used [\ion{O}{III}] $4959/5007\ \si{\angstrom}$ doublet), expanding the range of redshifts and instrument configurations for which precise flux-ratio measurements can be obtained.  Five other quadruply imaged quasars have already been observed with JWST/NIRSpec IFU for time-delay cosmography purposes \citep{TDCOSMO2025}, and this sample includes some suitable candidates for flux-ratio anomaly measurements. Cycle 5 program GO-9637 will obtain similar data for 11 more systems. In addition, \cite{Nierenberg2014} demonstrated that IFS with adaptive optics can also be used to precisely measure narrow-line flux ratios. The newly commissioned KAPA \citep[Keck All-sky Precision Adaptive-optics,][]{KAPA, Surendran2026} project, in particular, brings increased capabilities to the OSIRIS instrument on the Keck~I telescope, providing a powerful means to determine narrow-line flux ratios. The [\ion{S}{III}] doublet will be a particularly valuable addition to the arsenal of narrow-line features for OSIRIS, since it falls in the Kbb bandpass for systems with source redshifts $z_s \sim 1.3-1.4$ while other prevalent doublets are inaccessible. In the more distant future, the next generation of 30-m class telescopes should enable rapid and high-precision narrow-line measurements \citep{Zelko2024} for the hundreds of new quads expected to be discovered by the Vera C. Rubin Observatory \citep{Shajib2025}, and the \textit{Euclid} and Roman space telescopes \citep{OguriMarshall2010}~; these larger samples will greatly strengthen the DM constraints obtained with flux-ratio anomaly statistics \citep{Nierenberg2026}.

\vspace{-0.3cm}
\section*{Acknowledgments}

We thank Matthew Malkan, William Sheu, and Maria Perez Mendoza for helpful discussions. This work was supported by the National Science Foundation under grant AST-2205100, and by the Gordon and Betty Moore Foundation under grant No. 8548. DG acknowledges support provided by the Brinson Foundation through a Brinson Prize Fellowship grant. This work is based on observations made with the NASA/ESA/CSA James Webb Space Telescope. The data were obtained from the Mikulski Archive for Space Telescopes at the Space Telescope Science Institute, which is operated by the Association of Universities for Research in Astronomy, Inc., under NASA contract NAS 5-03127 for JWST. These observations are associated with program GO-1794 (PI: Suyu; co-PIs: Yıldırım, Treu).

\vspace{-0.2cm}
\section*{Data Availability}

The specific observations analyzed in this work are publicly available online and can be accessed at \url{https://dx.doi.org/10.17909/wjva-0750}. The software package \texttt{lensqso-specfit} that was used to model the spectra is openly available in the public code repository \url{https://github.com/Lens-Dark-Matter/lensqso-specfit}.

\vspace{-0.2cm}
\section*{Software}

This research made use of the following software packages: \texttt{astropy} \citep{astropy1, astropy2, astropy3}, \texttt{COBYQA} \citep{COBYQA1, COBYQA2}, \texttt{corner} \citep{corner}, \texttt{emcee} \citep{emcee}, \texttt{jupyter} \citep{jupyter}, \texttt{lenstronomy} \citep{Birrer2015, lenstronomy, lenstronomy2}, \texttt{matplotlib} \citep{matplotlib}, \texttt{numpy} \citep{numpy}, \texttt{pandas} \citep{pandas, pandas_repo}, \texttt{raccoon} \citep{raccoon}, \texttt{RegalJumper} \citep{Shajib2026},  \texttt{scipy} \citep{scipy}, and \texttt{stpsf} \citep{stpsf}.

\bibliographystyle{mnras}
\bibliography{biblio} 

\appendix
\section{Wiggle correction} 
\label{app:wiggle_correction}

In IFU mode, NIRSpec has a spatially undersampled PSF: at $3\mu$m for instance, its FWHM is comparable to the $0 \farcs 1$ native spaxel size \citep[e.g.,][]{Ruffio2024}. When the dither pattern does not allow to recover the optimal sampling, this creates resampling noise which manifests itself as low-frequency sinusoidal-like artifacts, or “wiggles”, in individual spaxels of the reduced 3D data cube \citep[e.g.,][]{Law2023,Dumont2025}. These wiggles can significantly distort the overall spectral shape and therefore bias line measurements.

To correct for these artifacts, we followed the procedure outlined in \cite{Shajib2026}, and used the publicly available software package \texttt{raccoon} \footnote{ \url{https://github.com/ajshajib/raccoon}} \citep{raccoon}. The cleaning strategy relies on the fact that wiggles impact single-spaxel spectra, but tend to average out when considering spectra summed over apertures with a size equal to or larger than the FWHM of the PSF, which corresponds to $\approx 3$ pixels in our case \citep{Shajib2026}.

The observed spectrum $d(\lambda)$ for an individual spaxel is fit with the following model:
\begin{equation}
	m (\lambda) = W(\lambda) \, T(\lambda),
\end{equation}
where $W(\lambda)$ is the wiggle signal, and $T(\lambda)$ is a template for the “true” underlying spectrum, constructed using the signal of neighboring spaxels:
\begin{equation}
    T(\lambda) = c_1\ a(\lambda) + c_2 \  s(\lambda) + c_3 \  \lambda^b + P(\lambda),
    \label{eq:template_spaxel_spectra}
\end{equation}
where $b,c_1, c_2, c_3$ are free parameters, $a(\lambda)$ is the spectrum summed over the spaxels in the circular aperture of radius $R_1$, $s(\lambda)$ is the spectrum summed over the spaxels in the annular aperture of inner radius $R_2$ and outer radius $R_3$. $P(\lambda)$ is a polynomial term of degree $d$ (we set $d=2$) with free coefficients, which, along with the power-law term $c_3\ \lambda^b$, is intended to account for differences in the continuum level between the aperture-summed and single-spaxel spectra. Similarly, the linear combination of $a(\lambda)$ and $s(\lambda)$ is designed to capture spatial variations in the shape of emission/absorption lines \citep{Shajib2026, Dumont2025}.

The wiggle signal itself is modeled as a sinusoidal chirp function
\begin{equation}
	W(\lambda) = 1 + A(\lambda) \sin \left[ \phi(\lambda) \right],
\end{equation}
where $A(\lambda)$ is the amplitude modulation function and $\phi(\lambda)$ is the wavelength-dependent phase, which are both modeled using cubic cardinal B-splines with $N_1$ and $N_2$ internal knots\footnote{If the wavelength range is divided into $N+2$ interval knots and modeled with a B-spline of degree $d$, the knot array needs to be extended with $d$ endpoints on each side (usually repeating the first and last knot) since a single B-spline already extends over $d+1$ knots. The number of \textit{internal} knots is then $N$, and there are $N+1$ intervals where the spline is non-zero.}, respectively. The uncertainty in the results of the wiggle model fit is estimated and propagated in the noise levels of the corrected spectrum $d(\lambda)/W(\lambda)$.

\begin{figure}
    \centering
    \includegraphics[width=0.95\columnwidth]{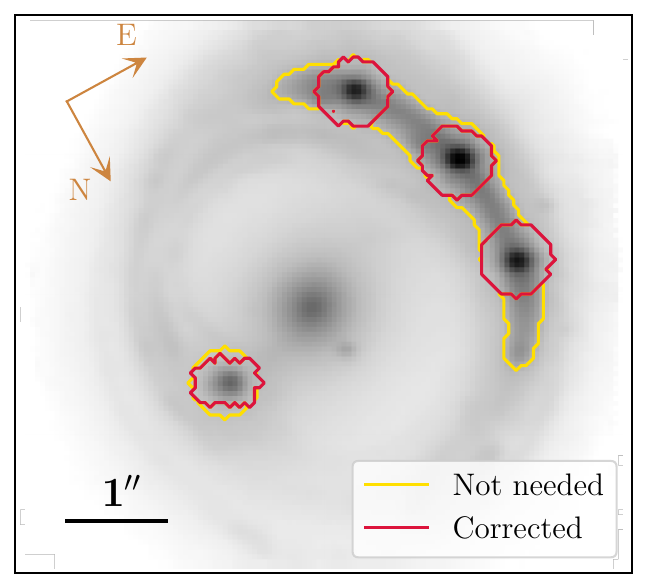}
    \caption{White-light image summed over the entire NIRSpec range, with regions showing the spaxels for which wiggle correction was considered but not needed (in yellow), and those which were effectively corrected (in red), i.e., where the wiggle signal was strong enough to be robustly modeled and removed (see text).}
    \label{fig:wiggle_cleaned_pix}
\end{figure} 

While \cite{Shajib2026} used the same datacube and already corrected the wiggles for the spaxels centered around the lensing galaxy, we are interested in extracting the spectra for the quasar images. We therefore attempted to clean spaxels that were within 7 pixels ($0\farcs35$ with the drizzled pixel size) of the quasar image locations, as well as spaxels in the lensed arc with a per-pixel signal-to-noise ratio ${\rm SNR}_{\rm pix} > 500$ (in order to accurately remove the extended emission at the location of the quasar). Among those spaxels, we only applied the corrections when the wiggle signal was detected with more than $3\sigma$ significance, and when the wiggle removed at least half of the variance observed in $ d(\lambda)/T(\lambda)$. This subset of effectively corrected spaxels, along with the initial selection, is displayed in Figure~\ref{fig:wiggle_cleaned_pix}. 

\begin{figure*}
    \centering
    \includegraphics[width=0.9\textwidth]{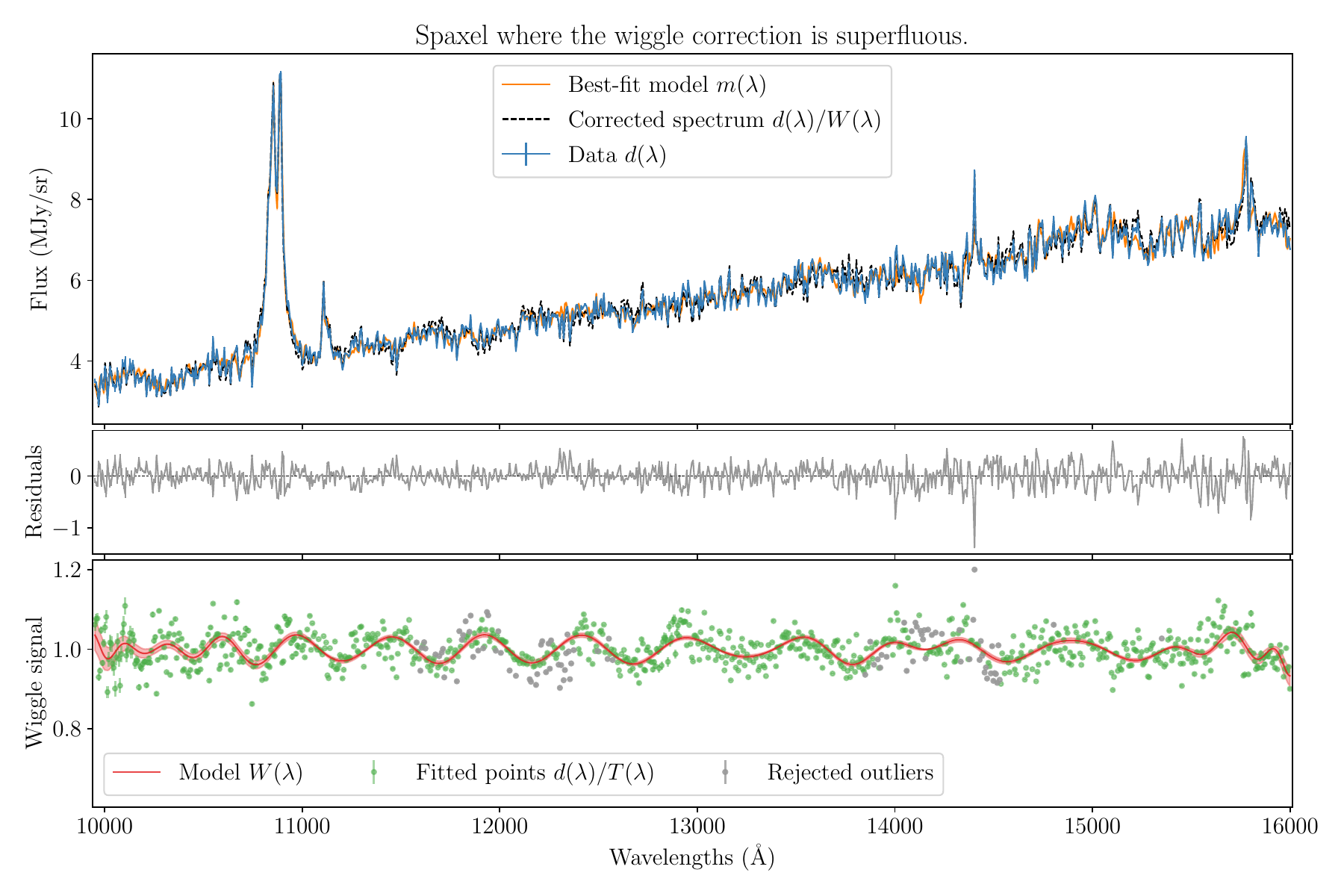}
    \includegraphics[width=0.9\textwidth]{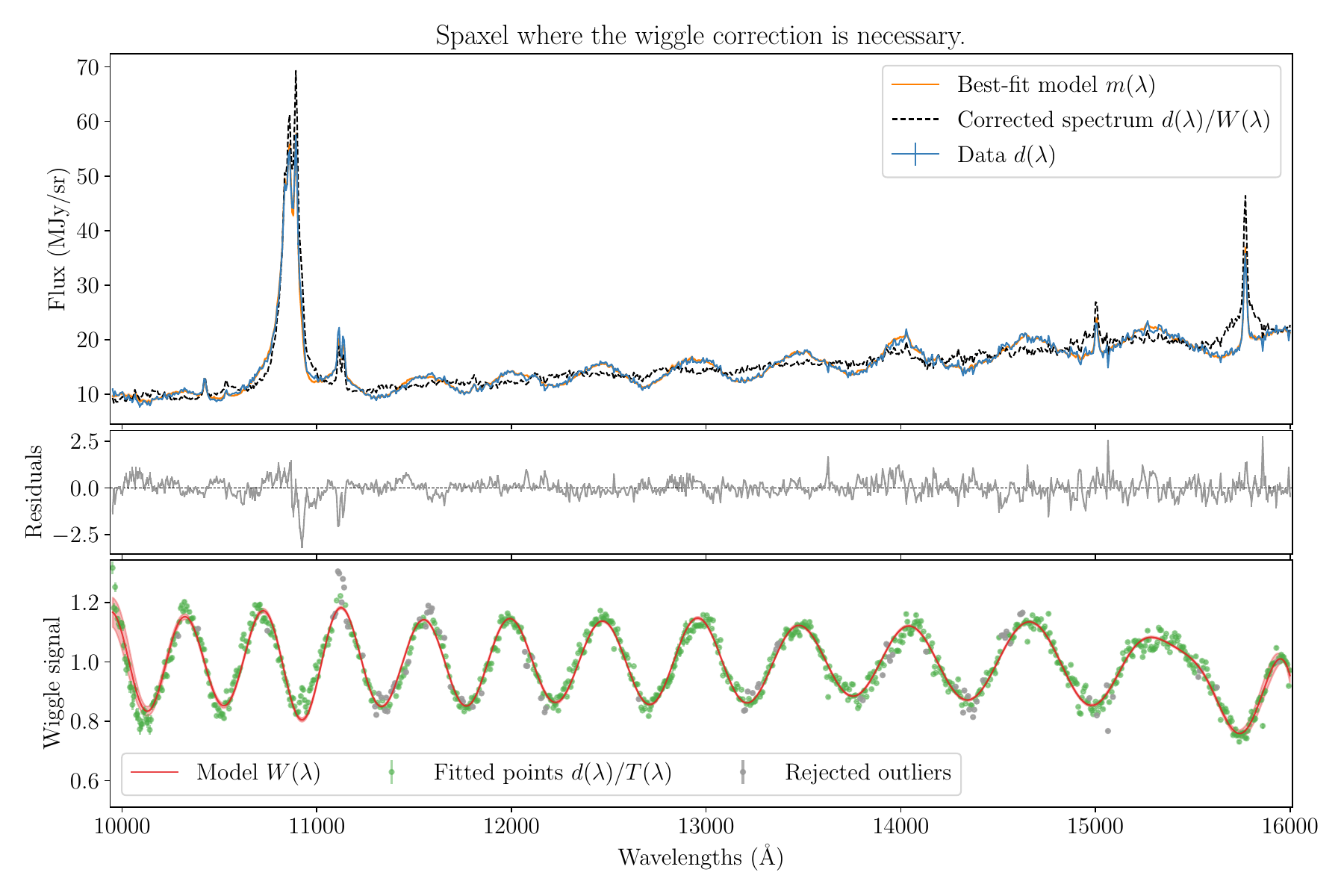}
    \caption{Two examples of single-spaxel spectrum cleaned using the wiggle correction procedure described in Appendix~\ref{app:wiggle_correction}, one near image D where the wiggle signal is inconsequential (top), and one near image A where it is prominent (bottom). For each spaxel, the top panel shows the uncorrected data, the best fit using the “template” spectrum and the wiggle model, and the corrected spectrum. The residuals are displayed in the middle panel, and the wiggle signal along with its sinusoidal chirp model in the bottom panel.}
    \label{fig:wiggle_correction_ex}
\end{figure*}

Even though the quasar spectra are only extracted and modeled in a narrow wavelength range (see section~\ref{sec:method}), we fit and correct the wiggles over the full range of the reduced data ($0.97-1.6\ \mu$m). This allows a better characterization of the global trends in the wiggle signal, while reducing the impact of noise and local outliers. We further mitigate this by applying an outlier rejection step (Benjamini-Hochberg procedure with a default false discovery rate control level $\alpha=0.05$). By default, we chose $R_1=3$ spaxels, $R_2=2$ spaxels, and $R_3=4$ spaxels for the aperture radii, and $N_1=10, N_2=3$ for the amplitude and phase modulations. We then visually inspected the wiggle model for each individual spaxel: for $\sim~85\%$, the best-fit model captured the signal faithfully with these default settings. For the remaining $\sim~15\%$ which had clear artifacts in the corrected spaxel spectrum, or unsatisfactory fit of the wiggle signal, we changed the settings until a satisfactory model was achieved, either by increasing the aperture sizes, increasing the number of knots for the B-splines, or decreasing the outlier rejection rate. We illustrate the procedure by displaying the final models for two example spaxels in Figure~\ref{fig:wiggle_correction_ex}, one where the wiggle signal is inconsequential, and one where it is prominent and correction is therefore necessary. 

\section{PSF uncertainties}
\label{app:PSF_uncertainties}

\begin{figure*}
    \centering
    \includegraphics[width=\textwidth]{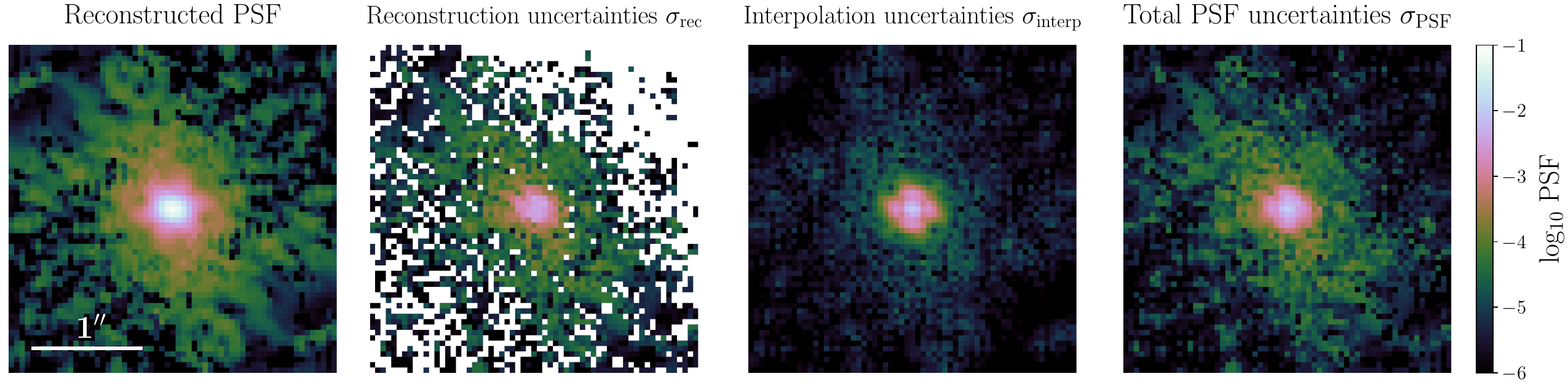}
    \caption{Illustration of the estimation of PSF uncertainties. \textbf{From left to right:} (i) Best estimate output by the PSF reconstruction procedure, (ii) Uncertainty map estimated by the PSF reconstruction procedure, (iii) Per-pixel interpolation uncertainties, estimated from the Hessian of the best-fit PSF according to Equation~\ref{eq:PSF_Hessian_unc},
    (iv) Total uncertainties, computed using the quadrature sum of (ii) and (iii).}
    \label{fig:PSF_unc}
\end{figure*}

We estimated the uncertainties on the reconstructed PSF using $\sigma^2_{\rm PSF}(i,j) = \sigma^2_{\rm rec}(i,j) + \sigma^2_{\rm interp}(i,j)$, i.e., as the combination in quadrature of two terms:
\begin{itemize}
    \item a PSF per-pixel variance map $\sigma^2_{\rm rec}(i,j)$ estimated during the PSF reconstruction step, and 
    \item an error term $\sigma^2_{\rm interp}(i,j)$ based on the second derivatives of the best PSF estimate, to account for interpolation uncertainties on the PSF grid, which we motivate below.
\end{itemize}
\vspace{-0.2cm}
We display both terms in Figure~\ref{fig:PSF_unc}, along with the best-fit PSF and total uncertainty map.

Let us consider a PSF model ${\rm PSF}(x,y)$ represented by a pixelized PSF grid $\{{\rm PSF}_{i,j}\}$ with pixel size $\delta_{\rm pix}$, and the pixelized image $\{I_{k,l}\}$ of a point source located at $(x_{\rm ps}, y_{\rm ps})$. The value at pixel $(k,l)$ is:
\begin{equation}
\begin{split}
    I_{k,l}  &= A_{\rm ps} \times{\rm PSF}(k-x_{\rm ps},l-y_{\rm ps}) \\
    &= A_{\rm ps} \times{\rm PSF}( i+\delta x, j+\delta y)
\end{split}
\end{equation}
where $A_{\rm ps} $ is the point source amplitude, $(i, j)$ are integer coordinates on the PSF grid, and $\delta \vec{x} = (\delta x, \delta y)$ are the offsets relative to the center of that pixel. Assuming that $\delta \vec{x}$ is small, we can perform the following Taylor expansion:
\begin{equation}
\begin{split}
    {\rm PSF}( i+\delta x, j+\delta y) = & \ {\rm PSF}_{i,j} \ + \delta \vec{x} \cdot \vec{\nabla} {\rm PSF} \bigr\rvert_{(i,j)}  \\ &  + \frac{1}{2} \delta \vec{x}^T \mathbf{H}[{\rm PSF}]\Bigr\rvert_{(i,j)} \delta \vec{x} + O(\|\delta \vec{x}\|^3)
\end{split}
\label{eq:Taylor_exp_PSF_interp}
\end{equation}
where the vertical bar denotes evaluation at specific coordinates, and 
\begin{equation}
    \mathbf{H}[{\rm PSF}] = \begin{bmatrix} \frac{\partial^2{\rm PSF}}{\partial x^2} & \frac{\partial^2{\rm PSF}}{\partial x \partial y} \\ \frac{\partial^2{\rm PSF}}{\partial x \partial y}  & \frac{\partial^2{\rm PSF}}{\partial y^2}  \end{bmatrix}
\end{equation}
is the Hessian matrix of the PSF. In \texttt{lenstronomy}, the point source contribution to the image pixels are computed using bilinear interpolation of the PSF grid: for $\delta x, \delta y \leq \delta_{\rm pix}$, this method will account for the first two terms (grid evaluation and gradient), plus the $\delta x \delta y$ term in the Hessian part of Equation~(\ref{eq:Taylor_exp_PSF_interp}), with derivatives estimated using finite differences. To leading order in $\|\delta \vec{x}\|$, the interpolation error in the PSF map due to pixel $(i,j)$ is therefore:
\begin{equation}
    \varepsilon_{i,j} = \frac{1}{2}\left( \frac{\partial^2{\rm PSF}}{\partial x^2}\bigr\rvert_{(i,j)} \delta x^2 + \frac{\partial^2{\rm PSF}}{\partial y^2} \bigr\rvert_{(i,j)} \delta y ^2 \right).
\end{equation}

We can then assume that the pixel offsets are uniformly distributed in the range where the PSF pixel will contribute to the  ``painting'' of the point source on the image, i.e., $\delta \vec{x}\sim\mathcal{U}([-\delta_{\rm pix}, \delta_{\rm pix}]^2)$. Thus, we have $\mathbb{E}[\delta x^2] = \mathbb{E}[\delta y^2] = \delta_{\rm pix}^2/3 $, $\mathbb{E}[\delta x^2\delta y^2]=\mathbb{E}[\delta x^2]\mathbb{E}[\delta y^2]$, and $\mathbb{E}[\delta x^4] = \mathbb{E}[\delta y^4] = \delta_{\rm pix}^4/5$, where $\mathbb{E}[.]$ is the expected value. Finally, we can estimate a per-pixel variance contribution from interpolation by calculating the expected mean square value of $\varepsilon_{i,j}$ in each pixel:

\begin{equation}
\begin{split}
    \sigma^2_{\rm interp}&(i,j) =  \mathbb{E}[\varepsilon_{i,j}^2] & \\
    &= \frac{\delta_{\rm pix}^4}{4} \Bigg[ \frac{1}{5} \left( \frac{\partial^2{\rm PSF}}{\partial x^2} \right)^2+ \frac{1}{5} \left( \frac{\partial^2{\rm PSF}}{\partial y^2} \right)^2 + \frac{2}{9} \frac{\partial^2{\rm PSF}}{\partial x^2} \frac{\partial^2{\rm PSF}}{\partial y^2} \Bigg]_{(i,j)}
\end{split}
\label{eq:PSF_Hessian_unc}
\end{equation}
(where we estimate derivatives on the grid using finite differences).

\section{Spectral fitting code}
\label{app:lensqso_specfit}

In this Appendix, we briefly describe some of the design choices made for the publicly available software package \texttt{lensqso-specfit}, including the treatment of model parameters by separating linear and nonlinear contributions (Appendix~\ref{app:lin_nonlin_params}), and the spectral features that are currently implemented (Appendix~\ref{app:spec_feat}).

\subsection{Optimization of model parameters}
\label{app:lin_nonlin_params}

To accelerate the spectral fitting procedure, we separate the model parameters into nonlinear and linear parameters, where the optimization for the second group can be posed as a weighted linear least square problem and therefore be solved for and marginalized over analytically. This is inspired by the treatment of parameters in \texttt{lenstronomy} \citep{Birrer2015}.

Let us consider an observed spectrum (i.e., flux measurements with uncertainties $\{ d_i \pm \sigma_i \}_{1\leq i \leq N}$ at some wavelengths $\{\lambda_i \}_{1\leq i \leq N}$). We will assume that the errors are Gaussian and the flux measurements independent. Let $\mathbf{D} \equiv [d_1, ..., d_N ]^T\in \mathbb{R}^N$ be the vector of data points and $\mathbf{W} = {\rm diag} (1/\sigma_1^2, ..., 1/\sigma_N^2)$ the inverse of the  covariance matrix (of dimension $N\times N$). Given a parametrized model described by parameter vector $\mathbf{q} \in \mathbb{R}^n$, we can decompose $\mathbf{q} = (\mathbf{X}, \boldsymbol{\theta})$ into linear parameters $\mathbf{X} \in \mathbb{R}^m$ and nonlinear parameters $\boldsymbol{\theta} \in \mathbb{R}^{n-m}$, where the linear parameters are defined by the existence of a matrix $\mathbf{M}(\boldsymbol{\theta})$ of dimension $N\times m$ such that $ \forall \mathbf{X}, \ \mathbf{S}(\mathbf{q}) = \mathbf{M}(\boldsymbol{\theta}) \cdot \mathbf{X} $ is the model spectrum. For instance, if the model profile is a Gaussian, the amplitude is a linear parameter, while the mean and width are nonlinear. For each spectral feature available to this date in \texttt{lensqso-specfit} (see Appendix~\ref{app:spec_feat}), we give the decomposition into linear and nonlinear parameters in Table~\ref{tab:lin_nonlin_params}. We note that the linear parameters of a single feature will also be linear in a multi-component model, since the fluxes from each feature are summed.

Our objective is to determine the posterior distribution of all the model parameters $\mathrm{p}( \mathbf{q}| \mathbf{D}) \propto \mathrm{p}( \mathbf{D}| \mathbf{q}) \cdot \mathrm{p}(\mathbf{q})$. The log-likelihood is
\begin{equation}
    \log\mathrm{p}( \mathbf{D}| \mathbf{q}) = -\left\|\mathbf{W}^{1/2}\cdot \left(\mathbf{D}-\mathbf{S}(\mathbf{q})\right) \right\|_2 
    \label{eq:log_lik_def}
\end{equation}
where $\|.\|_2$ is the Euclidian norm. For a given choice of nonlinear parameters, optimizing the linear parameters thus reduces to a weighted linear least square problem: the maximum-likelihood estimator is
\begin{equation}
    \hat{\mathbf{X}}(\boldsymbol{\theta})  = \left( \mathbf{M}(\boldsymbol{\theta})^T \mathbf{W} \mathbf{M}(\boldsymbol{\theta})\right)^{-1} \cdot \mathbf{M}(\boldsymbol{\theta})^T \mathbf{W}\mathbf{D}
\end{equation}
and the estimated variance-covariance matrix of $\hat{\mathbf{X}}$ is
\begin{equation}
    \boldsymbol{\Sigma}_{X}(\boldsymbol{\theta}) = \left( \mathbf{M}(\boldsymbol{\theta})^T \mathbf{W} \mathbf{M}(\boldsymbol{\theta})\right)^{-1},
\end{equation}
such that, in the vicinity of $\hat{\mathbf{X}}$, we can approximate the log-likelihood with the following second-order Taylor expansion:
\begin{equation}
\label{eq:likelihood_approx}
\begin{split}
     \log\mathrm{p}( \mathbf{D}| \mathbf{q}) &=  \log\mathrm{p}( \mathbf{D}| \mathbf{X},\boldsymbol{\theta}) \\
     &\approx \log\mathrm{p}( \mathbf{D}| \hat{\mathbf{X}}(\boldsymbol{\theta}),\boldsymbol{\theta}) -\frac{1}{2} (\mathbf{X}-\hat{\mathbf{X}}(\boldsymbol{\theta}))^T \boldsymbol{\Sigma}_{X}(\boldsymbol{\theta})^{-1} (\mathbf{X}-\hat{\mathbf{X}}(\boldsymbol{\theta})).
\end{split}
\end{equation}
The likelihood of the data given nonlinear parameters $\boldsymbol{\theta}$ can then be written as a marginalization over the linear parameters
\begin{equation}
    \mathrm{p}( \mathbf{D}| \boldsymbol{\theta}) = \int d\mathbf{X} \ \mathrm{p}( \mathbf{D}| \mathbf{X},\boldsymbol{\theta}) \mathrm{p}(\mathbf{X}| \boldsymbol{\theta})
\end{equation}
where $\mathrm{p}(\mathbf{X}| \boldsymbol{\theta})$ can be treated a constant if we assume a prior on the linear parameters that is independent of the nonlinear parameters, and uniform over a sufficiently large volume of parameter space. Then, using the approximation from Eq.~(\ref{eq:likelihood_approx}), we obtain
\begin{equation}
    \mathrm{p}( \mathbf{D}| \boldsymbol{\theta}) \propto (2\pi)^{m/2} \sqrt{|\det(\boldsymbol{\Sigma}_{X}(\theta))|}\ \mathrm{p}( \mathbf{D}|  \hat{\mathbf{X}}(\boldsymbol{\theta}),\boldsymbol{\theta}).
    \label{eq:approx_gaussian_lik}
\end{equation}
where $\mathrm{p}( \mathbf{D}|  \hat{\mathbf{X}}(\boldsymbol{\theta}),\boldsymbol{\theta})$ is evaluated using Eq.~(\ref{eq:log_lik_def}). This shows that the likelihood marginalized over linear parameters can be calculated with relatively inexpensive operations (e.g., matrix inversion), thus, only the space of nonlinear parameter needs to be explored with PSO, constrained optimization or MCMC sampling, and this reduction of dimensionality accelerates the fitting process.

We note that this approach can be easily generalized to the simultaneous fitting of multiple spectra $\mathbf{D}_1, ...,\mathbf{D}_p$ with shared parameters, e.g., in the case of a multiply imaged quasar. We can indeed decompose the nonlinear parameters in the model as $\boldsymbol{\theta} = (\boldsymbol{\theta}_1, ..., \boldsymbol{\theta}_p, \boldsymbol{\theta}_{\rm sh})$ where $\boldsymbol{\theta}_l$ are the parameters only relevant for $\mathbf{D}_l$, and $\boldsymbol{\theta}_{\rm sh}$ are the shared parameters. Then, we can write the joint likelihood as
\begin{equation}
    p( \mathbf{D}_1, ...,\mathbf{D}_p | \boldsymbol{\theta} ) = \prod_{l=1}^p p(\mathbf{D}_l | \boldsymbol{\theta}_l, \boldsymbol{\theta}_{\rm sh})
    \label{eq:joint_lik_spectra}
\end{equation}
and evaluate each term of the product using Equation~(\ref{eq:approx_gaussian_lik}) with the data $\mathbf{D}_l $ and the set of nonlinear parameters $\boldsymbol{\tilde{\theta}}_l = ( \boldsymbol{\theta}_l, \boldsymbol{\theta}_{\rm sh})$. To write Equation~(\ref{eq:joint_lik_spectra}), we implicitly assumed that the linear parameters $\mathbf{X}_l$ for each spectrum are independent (such that the $\mathbf{D}_l $ are conditionally independent given $\boldsymbol{\theta}$), which means that only nonlinear parameters can be shared between the spectral models. For simplicity, we keep simple notations like $p(\mathbf{D}|\boldsymbol{\theta})$ in the rest of our discussion.

The posterior on nonlinear parameters is 
$\mathrm{p}( \boldsymbol{\theta} | \mathbf{D}) \propto  \mathrm{p}( \mathbf{D}| \boldsymbol{\theta}) \cdot  \mathrm{p}( \boldsymbol{\theta})$, and we choose a prior distribution
\begin{equation}
    \mathrm{p}( \boldsymbol{\theta}) \propto \pi( \boldsymbol{\theta}) \cdot\prod_j\mathbbm{1}\left( \min_{1\leq i \leq N} [ \mathbf{S}_j ( \hat{\mathbf{X}}(\boldsymbol{\theta}),\boldsymbol{\theta})]_i \geq \varepsilon\right)
    \label{eq:prior_spec}
\end{equation}
where $\pi( \boldsymbol{\theta})$ represents user-inputted uniform priors, $\mathbbm{1}$ is the indicator function, $\mathbf{S}_j \in \mathbb{R}^N$ is the vector of flux contributions (at all the $\lambda_i$) from the $j$th component of the spectral model, and $\varepsilon<0$ is a numerical tolerance. 
The first term allows to set allowed ranges for the nonlinear parameters: for instance, for a Gauss-Hermite series of width $\sigma$ (see Appendix~\ref{app:line_profiles}), we set default priors on the FHWM using $1000 \ \text{km/s} < 2\sqrt{2\ln2}\sigma < 10000 \ \text{km/s} $ for broad lines and $100 \ \text{km/s} < 2\sqrt{2\ln2}\sigma < 1000 \ \text{km/s} $ for narrow lines. The second term of Equation~(\ref{eq:prior_spec}) is here to ensure that individual emission features do not go to large negative flux values, which would allow unphysical fits to the data. The numerical tolerance enables faster convergence toward physical solutions, in particular for lines modeled with Gauss-Hermite series, which can have small negative values in the wings. We remark that, due to this near-positivity check, the current implementation does not allow for absorption features, but this could be easily realized by replacing the corresponding term in the product by $\mathbbm{1}\big( \max_{1\leq i \leq N} [ \mathbf{S}_j ( \hat{\mathbf{X}}(\boldsymbol{\theta}),\boldsymbol{\theta})]_i \leq -\varepsilon\big) $.

Following the formalism described above, the MCMC sampler in \texttt{lensqso-specfit} outputs a posterior on the nonlinear parameters, marginalized over the linear parameters. We sometimes need to consider the joint posterior, however, which we can write as 
\begin{equation}
     \mathrm{p}( \mathbf{X}, \boldsymbol{\theta} | \mathbf{D}) =  \mathrm{p}( \mathbf{X} | \mathbf{D}, \boldsymbol{\theta}) \ \mathrm{p}( \boldsymbol{\theta} | \mathbf{D}).
     \label{eq:joint_posterior_lin_nonlin}
\end{equation}
where we know that $\mathrm{p}( \mathbf{X} | \mathbf{D}, \boldsymbol{\theta})$ can be approximated as a multivariate normal distribution with mean $\hat{\mathbf{X}}(\boldsymbol{\theta})$ and covariance matrix $ \boldsymbol{\Sigma}_{X}(\boldsymbol{\theta})$.
For instance, the narrow-line flux ratios $\{f_i/f_j\}$ (here, viewed as a vector with arbitrary indexing) depend on the amplitude of the narrow doublets (which are linear parameters), and sometimes on the (nonlinear) shape parameters ($c_i/c_0$ in the case of a Gauss-Hermite series, see Equation~(\ref{eq:GaussHermite_integral})), i.e., there exists a function $f$ such that $\{f_i/f_j\} = f(\mathbf{X}, \boldsymbol{\theta} )$. Equation~(\ref{eq:joint_posterior_lin_nonlin}) implies that, to draw samples from the flux-ratio posterior $\mathrm{p}( \{f_i/f_j\} | \mathbf{D})$, we simply need to (i) draw samples $\boldsymbol{\theta}_k$ from $\mathrm{p}( \boldsymbol{\theta}_k | \mathbf{D})$, i.e., from the MCMC chains, then (ii) for each $\boldsymbol{\theta}_k$, draw $X_k$ from $\mathcal{N}( \hat{\mathbf{X}}(\boldsymbol{\theta_k}),  \boldsymbol{\Sigma}_{X}(\boldsymbol{\theta_k}))$, and finally (iii) calculate $f(X_k, \boldsymbol{\theta}_k )$.

\defcitealias{Kovacevic2010}{K10}
\defcitealias{Garcia-Rissmann2012}{GR12}

\setlength\tabcolsep{3pt} 
\setlength{\extrarowheight}{3pt}
\begin{table}
\caption{\label{tab:lin_nonlin_params} Partition into linear and nonlinear parameters for each of the spectral features available to date in \texttt{lensqso-specfit}. NB: For the doublets, the two lines share the same parameters (linear and nonlinear) but the flux of the second line is multiplied by a fixed relative amplitude afterward.}
\centering
\begin{tabular}{cccc}
\hline &
\textrm{Linear}&
\textrm{Nonlinear} & \textrm{Equation} \vspace{-0.2cm}
 \\
\textrm{Model profile}&
\textrm{parameters $\mathbf{X}$}&
\textrm{parameters $\boldsymbol{\theta}$} & \textrm{or Reference} \\
  \hline  \hline 
Power-law & $A_c$ & $\beta$ & (\ref{eq:power-law_continuum})  \\
Polynomial & $\{p_i\}_{i\geq0}$ & - & (\ref{eq:polynomial_continuum}) \\
  \hline  
Gauss-Hermite series $^\dagger$ & $\{c_i\}_{i\geq0}$  & $\Delta\lambda, \sigma$ & (\ref{eq:GaussHermite_series}) \\
 & $c_0$ & $\Delta\lambda, \sigma, \left\{\frac{c_i}{c_0}\right\}_{i\geq1}$ & (\ref{eq:GaussHermite_series}) \\
Voigt & $A$ & $\Delta\lambda, \sigma, \gamma$  & (\ref{eq:voigt_profile}) 
  \\ \hline 
Optical \ion{Fe}{II} template $^\dagger$ & $F, G, S, {\rm IZw1}$ & $W, d$  & \citetalias{Kovacevic2010} \\
 & $F$ & $W, d, \frac{G}{F}, \frac{{\rm IZw1}}{F}, \frac{S}{F}$  & \\
NIR \ion{Fe}{II}\,+\,\ion{Mg}{II} template & $A_{9997}$ & $\Delta\lambda, \sigma$ & \citetalias{Garcia-Rissmann2012}
 \\
  \hline  \hline
\end{tabular}\\
\footnotesize{$^\dagger$ Depending on modeling choices, coefficients for the Gauss-Hermite series and the optical \ion{Fe}{II} template (see Appendix~\ref{app:spec_feat}) can either be treated as linear or nonlinear parameters.}
\end{table}

\subsection{Available spectral features}
\label{app:spec_feat}

\subsubsection{Line profiles}
\label{app:line_profiles}

\paragraph*{Gauss-Hermite series} We define the Gauss-Hermite functions with the following convention:
\begin{equation}
\psi_n(x) = (2^n \ n!\sqrt{\pi}\ )^{-1/2} H_n(x)\, e^{-x^2/2},
\end{equation}
where $H_n(x) = (-1)^n e^{x^2} \frac{d^n}{dx^n} e^{-x^2}$ for $n\in \mathbb{N}$ are the Hermite polynomials defined in the physicist's convention. In particular, the first four polynomials are given by $H_0(x) = 1$, $H_1(x) = 2x$, $H_2(x) = 4x^2 - 2$, and $H_3(x) = 8x^3 - 12x$. With this definition, we have 
\begin{equation}
    \int_{-\infty}^{\infty} \psi_n(x) \psi_m(x)dx = \delta_{mn}
    \label{eq:GaussHermite_integral}
\end{equation}
where $\delta_{ij}$ is the Kronecker delta symbol. In fact, the $\{\psi_n\}$ form an orthonormal basis of $L^2(\mathbb{R})$. A spectral line profile centered at $\lambda_0 = \lambda_{\rm vac} + \Delta\lambda$ (where $\lambda_{\rm vac} $ is the rest-frame vacuum wavelength) with characteristic width $\sigma$ can therefore be expanded in terms of these functions \citep[e.g.,][]{Gerhard1993, vanderMarelFranx1993, CappellariEmsellem2004}: to order $N$,
\begin{equation}
    \psi_{\rm GH}(\lambda) = \frac{\pi^{-1/4}}{\sqrt{2}\sigma}\sum_{n=0}^{N} c_n\, \psi_n\left(\frac{\lambda - \lambda_0}{\sigma}\right),
    \label{eq:GaussHermite_series}
\end{equation}
where $c_0, ..., c_N$ are free coefficients describing the amplitude and shape of the line. We can write
\begin{equation}
\begin{split}
    \int_{-\infty}^{\infty} \psi_{\rm n}(x)dx &= (2^n \ n!\sqrt{\pi}\ )^{-1/2} \int_{-\infty}^{\infty}  H_n(x)\, e^{-x^2/2}dx \\
    &= (2^n \ n!\sqrt{\pi}\ )^{-1/2}H_n(0) \sqrt{2\pi}\\
    &= \begin{cases}
        0 \quad \text{ if } n \text{ is odd} \\
        \frac{(n-1)!!}{\sqrt{n!}} \sqrt{2}\pi^{1/4} \quad \text{ if } n \text{ is even},
    \end{cases}
\end{split}
\end{equation}
where $!!$ denotes the semifactorial (i.e., $(n-1)!! = \prod_{k=1}^{n/2}(2k-1)$ for $n$ even, and $(-1)!!=1$), and we have used Equation~(7.376.1) from \cite{GradshteynRyzhik2007} to evaluate the integral. Then, with a change of variables $x=(\lambda - \lambda_0)/\sigma$, integrating the line profile yields
\begin{equation}
\begin{split}
     \int_{-\infty}^{\infty} \psi_{\rm GH}(\lambda) \,d\lambda &= \frac{\pi^{-1/4}}{\sqrt{2}}\sum_{n=0}^{N} c_n\ \int_{-\infty}^{\infty} \psi_{\rm n}(x)dx =  \sum_{\substack{n=0 \\ n \text{ even}}}^{N} c_n \frac{(n-1)!!}{\sqrt{n!}} \\
     &= c_0 + \frac{c_2}{\sqrt{2}} + \frac{c_4 \sqrt{6}}{4} + \frac{c_6\sqrt{5}}{4} +...
\end{split}
\end{equation}

We note that, in Equation~(\ref{eq:GaussHermite_series}), the term in $n=0$ corresponds to a Gaussian line profile with amplitude $c_0$ (in particular, to model a line with a simple Gaussian, we only need to set $N=0$), such that higher orders in the series can be interpreted as departures from Gaussianity—in particular $c_3$ imparts some asymmetric deviations (blue/red wings), while $c_4$ creates profiles that are more peaked ($c_4>0$) or flat-topped ($c_4<0$) than a Gaussian \citep[e.g.,][]{Riffel2010}. In practice, the first and second order coefficient in the series are degenerate with the properties (width, mean and amplitude) of the reference Gaussian \citep[e.g.,][]{vanderMarelFranx1993}, so we can fix $c_1=c_2=0$. Then, the standard practice is to use $N=4$ and to fit for $c_0$, $c_3$, and $ c_4$ (which is what we assumed for the spectral model presented in Section~\ref{subsec:spectral_model}). 

Furthermore, in \texttt{lensqso-specfit}, the coefficients of the Hermite series are either:
\begin{itemize}
    \item all treated as linear parameters (see Appendix~\ref{app:lin_nonlin_params}), in which case they cannot be shared between different quasar images, or
    \item  only $c_0$ is treated as a linear parameter, then the $c_i/c_0$ for $i\geq1$ are optimized as nonlinear parameters, such that the shape of the line profile can be shared between different quasar images, with $c_0$ acting as an amplitude scaling independently for each image.
\end{itemize} 
\vspace{-0.2cm}
Both options were used in Section~\ref{subsec:spectral_model}, since the shapes of the line profiles for [\ion{S}{III}], Pa\,$\varepsilon$, and Pa\,$\eta$ were assumed to be the same for all four images, while they were fitted independently for Pa\,$\zeta$.
\vspace{0.2cm}

Finally, we note that, for $n\geq1$, $\psi_n(x) <0$  for some values $x\in \mathbb{R}$, which can lead to negative flux values in $\psi_{\rm GH}(\lambda)$, depending on the values for the coefficients $\{c_n\}$. Since we only consider emission lines in our spectral models, negative flux values would be unphysical, but we still want to give some flexibility to the fitting procedure, which motivates the use for a numerical tolerance $\varepsilon <0$, that is, a negative value above which fluxes are still considered positive (see Section~\ref{app:lin_nonlin_params}).

\paragraph*{Voigt profile} The Voigt profile is a line shape resulting from the convolution of a Lorentzian and a Gaussian profile. If $\sigma$ is the standard deviation of the Gaussian and $\gamma$ the half-width at half-maximum of the Lorentzian, a Voigt profile centered at $\lambda_0 = \lambda_{\rm vac} + \Delta\lambda$ is 
\begin{equation}
    \psi_V(\lambda) = \frac{A}{\sigma\sqrt{2\pi}} \mathrm{Re}\left[w\left( \frac{(\lambda - \lambda_0) + i\gamma}{\sigma\sqrt{2}}\right)\right]
    \label{eq:voigt_profile}
\end{equation}
where $\mathrm{Re}[w(z)]$ is the real part of the Faddeeva function.
The profile is normalized such that $\int_{-\infty}^{\infty} \psi_V(\lambda)= A$, so the amplitude parameter $A$ directly represents the total integrated line flux. 
The FWHM of this profile can be approximated \citep{Whiting1968} as
\begin{equation}
    {\rm FWHM }(\psi_V) \approx \gamma + \sqrt{\gamma^2 + 8\ln(2)\sigma^2}
\end{equation}
Since the Voigt profile reduces to a Gaussian for $\gamma \ll \sigma$ and to a Lorentzian for $\gamma  \gg \sigma$, its use is relevant only for the case when $\gamma \sim 2\sqrt{2\ln(2)}\sigma $, in which case we can write 
\begin{equation}
    \sigma \sim \frac{{\rm FWHM}(\psi_V)}{2(\sqrt{2}+1)\sqrt{2\ln(2)}}, \quad \gamma \sim \frac{{\rm FWHM}(\psi_V)}{\sqrt{2}+1}.
\end{equation}
We use these relations to set the initial values and priors on model parameters when they not specified by the user: for instance, the default (uniform) prior is $1000 \ \text{km/s} <{\rm FWHM} < 10000 \ \text{km/s} $ for broad lines and $100 \ \text{km/s} <{\rm FWHM} < 1000 \ \text{km/s} $ for narrow lines.

\subsubsection{Other features: continuum and template lines}

\paragraph*{Continuum.} The continuum can either be modeled as a power law
\begin{equation}
    C(\lambda) = A_c \left( \frac{\lambda}{\lambda_c} \right)^\beta,
    \label{eq:power-law_continuum}
\end{equation}
where $\lambda_c$ is a fixed value (by default, the mean wavelength of the range considered for fitting) and $A_c, \beta$ are the amplitude and power-law index parameters, respectively ; or as an order $N$ polynomial (where $N$ is specified by the user)

\begin{equation}
    C(\lambda) = \sum_{n=0}^{N} p_n \left(\frac{\lambda - \lambda_c}{L}\right)^n,
    \label{eq:polynomial_continuum}
\end{equation}
where $\lambda_c, L$ are fixed values representing a reference wavelength and a typical scale for variations (by default, the mean wavelength and half-span of the fitted range), and $p_0,..., p_N$ are free coefficients.

\paragraph*{Optical \ion{Fe}{II} template.}
While irrelevant to the data presented in this work, a template for optical \ion{Fe}{II} emission is included in the software package in anticipation of future IFU analyses, since it can be important for the region around the [\ion{O}{III}] $4959/5007 \si{\angstrom}$ doublet, which is the most prominent narrow feature for many lensed quasar systems observed with JWST/NIRSpec or Keck/OSIRIS. Following \cite{Nierenberg2020}, we included the semi-empirical template presented in \cite{Kovacevic2010} (\citetalias{Kovacevic2010} in Table~\ref{tab:lin_nonlin_params}), which sorts the strongest \ion{Fe}{II} transitions in the $4400-5500\ \si{\angstrom}$ range into three theoretical groups of lines, and one empirical group observed in the spectrum of quasar IZw1. All the lines are represented as Gaussians with a common width $W$ and shift $d$, and the relative amplitudes within a group are fixed such that the remaining free parameters are the overall scalings ($F, G, S, {\rm IZw1}$) for each group. Similar to the Gauss-Hermite series, these can all be treated as linear parameters, or the amplitudes relative to group $F$ are treated as nonlinear to allow them to be shared among multiple images (see Table~\ref{tab:lin_nonlin_params}).

\paragraph*{NIR \ion{Fe}{II}\,+\,\ion{Mg}{II} template.} Even if it did not ultimately improve the spectral model for RXJ1131, we implemented the semi-empirical template for \ion{Fe}{II} and \ion{Mg}{II}  emission in the NIR ($8300-11600\ \si{\angstrom}$) from \cite{Garcia-Rissmann2012} (\citetalias{Garcia-Rissmann2012} in Table~\ref{tab:lin_nonlin_params}). This template is composed of lines identified by the theoretical work of \cite{SigutPradhan2003}, which have intensities relative to the $9997 \si{\angstrom}$ line that are derived from the observed spectrum of IZw1 (or, for some of the most prominent lines, from the best-fitting theoretical model). Assuming that all the lines can be represented by Gaussians \citep{Marinello2016}, the entire template is described by their common shift $\Delta\lambda$ and width $\sigma$, and by the amplitude $A_{9997}$ of the reference line.

\section{Alternative spectral models}
\label{app:FRs_alt_specmodels}

To assess the robustness of our spectral model, we explored a broad range of variations around the fiducial assumptions described in Section~\ref{subsec:spectral_model}. Table~\ref{tab:alt_spec_models} presents the [\ion{S}{III}] flux ratios relative to image A for a representative selection of these alternative models, spanning different choices for (i) line profiles (ii) shared versus image-specific nonlinear parameters, and (iii) the inclusion or exclusion of the faint line at $\sim 8990 \si{\angstrom}$ and of the \ion{Fe}{II}+\ion{Mg}{II} emission template. We also attempted to change the value for the theoretical [\ion{S}{III}] doublet ratio.

Given the large number of possibilities, we did not exhaust the full space of model combinations; however, other combinations with a large number of degrees of freedom (e.g., when using $N=4$ Gauss-Hermite profiles plus additional spectral features plus fully independent parameters across all images) tend to suffer from degeneracies, preventing convergence toward physically meaningful solutions (for instance, requiring large numerical tolerances to accommodate negative component fluxes). In Table~\ref{tab:alt_spec_models}, we only selected models that avoid this issue—none of these alternative configurations meaningfully improve the fit to the extracted spectral data, and the [\ion{S}{III}] flux-ratio values are all consistent with our fiducial measurements (presented in the first line of Table~\ref{tab:alt_spec_models}) within the estimated uncertainties. We thus conclude that the reported flux-ratio values are robust to spectral modeling assumptions.

\begin{table*}
\caption{Measured [\ion{S}{III}] flux ratios relative to image A for a variety of spectral modeling assumptions (line profiles, whether nonlinear parameters are shared between images or fit independently for each image, additional emission features). The fiducial model presented in Section~\ref{subsec:spectral_model} corresponds to the first line.}
\label{tab:alt_spec_models}
\centering
\begin{tabular}{c|cc|cc|cc|c|c|c|c|c}
\hline  & \multicolumn{2}{c|}{\textrm{Pa\,$\varepsilon$}}  & \multicolumn{2}{c|}{\textrm{Pa\,$\zeta$}}  & \multicolumn{2}{c|}{\textrm{Pa\,$\eta$}}  & \textrm{Line $^\dagger$ at} & \textrm{\ion{Fe}{II}+\ion{Mg}{II} } & \multicolumn{3}{c}{\textrm{[\ion{S}{III}] flux ratios}}  \\ \cline{2-3} \cline{4-5} \cline{6-7} \cline{10-12}
\textrm{[\ion{S}{III}] profile $^*$} &  \textrm{Profile $^*$} & \textrm{Shared ?} & \textrm{Profile $^*$} & \textrm{Shared ?} &  \textrm{Profile $^*$} & \textrm{Shared ?} & \ \textrm{$\sim 8990 \si{\angstrom}$} \
& \textrm{template} & $f_B/f_A$  & $f_C/f_A$  & $f_D/f_A$   \\
  \hline  \hline 
\textrm{GH ($N=4$)}  & \textrm{GH ($N=4$)}  & \cmark & \textrm{GH ($N=4$)}  & \xmark & \textrm{GH ($N=4$)}  & \cmark & \cmark & \xmark & $0.45 \pm 0.02$ &  $0.35 \pm 0.02$ & $0.0070 \pm 0.004$ \\
\textrm{GH ($N=4$)}  & \textrm{GH ($N=4$)}  & \cmark & \textrm{GH ($N=4$)}  & \xmark & \textrm{GH ($N=4$)}  & \cmark & \xmark & \xmark & $0.45 \pm 0.02$ & $0.35 \pm 0.02$ & $0.0070 \pm 0.004$  \\
\textrm{GH ($N=4$)}  & \textrm{GH ($N=4$)}  & \xmark & \textrm{GH ($N=4$)}  & \xmark & \textrm{GH ($N=4$)}  & \xmark & \cmark & \xmark & $0.43 \pm 0.02$ & $0.36 \pm 0.02$ & $0.0071 \pm 0.004$ \\
\textrm{GH ($N=3$)}  & \textrm{GH ($N=3$)}  & \xmark & \textrm{GH ($N=3$)}  & \xmark & \textrm{Gaussian}  & \xmark & \cmark & \xmark & $0.48 \pm 0.03$ &  $0.32 \pm 0.02$ & $0.0070 \pm 0.006$ \\
\textrm{Gaussian}  & \textrm{Gaussian}  & \xmark & \textrm{Gaussian}  & \xmark & \textrm{Gaussian}  & \xmark & \cmark & \xmark & $0.46 \pm 0.02$ &  $0.35 \pm 0.02$ & $0.0071 \pm 0.004$ \\
\textrm{Gaussian}  & \textrm{Gaussian}  & \xmark & \textrm{Gaussian}  & \xmark & \textrm{Gaussian}  & \xmark & \cmark & \cmark & $0.46 \pm 0.02$ &  $0.36 \pm 0.02$ & $0.0073 \pm 0.004$ \\
\textrm{Gaussian}  & \textrm{Gaussian}  & \xmark & \textrm{Gaussian}  & \xmark & \textrm{Gaussian}  & \xmark & \xmark & \cmark &  $0.46 \pm 0.02$ & $0.36 \pm 0.02$ & $0.0070 \pm 0.004$ \\
\textrm{Gaussian}  & \textrm{Gaussian}  & \xmark & \textrm{Gaussian}  & \xmark & \textrm{Gaussian}  & \xmark & \xmark & \xmark & $0.46 \pm 0.02$ & $0.35 \pm 0.02$ & $0.0071 \pm 0.004$ \\
\textrm{Voigt}  & \textrm{Voigt}  & \xmark & \textrm{Voigt}  & \xmark & \textrm{Voigt}  & \xmark & \cmark & \xmark & $0.46 \pm 0.02$ & $0.36 \pm 0.02$ & $0.0071 \pm 0.004$ \\
\textrm{GH ($N=4$)}  & \textrm{GH ($N=4$)}  & \cmark & \textrm{GH ($N=4$)}  & \xmark & \textrm{GH ($N=4$)}  & \cmark & \cmark & \xmark & $0.45 \pm 0.02$ & $0.34 \pm 0.02$ & $0.0072 \pm 0.004$ \\[-4pt]
\textrm{with } $R=2.59$ $^\ddag$ & & & & & & & & & & & \\
\hline  \hline
\end{tabular}\\
\vspace{0.1cm}
\footnotesize{$^*$ We write GH ($N=...$) for a Gauss-Hermite series of order $N$ with the 1st and 2nd order coefficients fixed to zero. \\
$^\dagger$ We attributed this faint line to \ion{N}{II} emission in Section~\ref{subsec:spectral_model}.\\
$^\ddag$ The [\ion{S}{III}] doublet ratio $A(\lambda9533)/ A(\lambda9071)$ is fixed at a theoretical value of $R=2.48$ \citep{Revalski2024} for all the other models. The value of $R=2.59$ is obtained by taking the ratio of transition probabilities listed in \cite{Podobedova2009}.}
\end{table*}

\vspace{-0.1cm}
\section{Correlation between flux ratios}
\label{app:correlation_FRs}
\begin{figure*}
    \centering
    \includegraphics[width=0.8\textwidth]{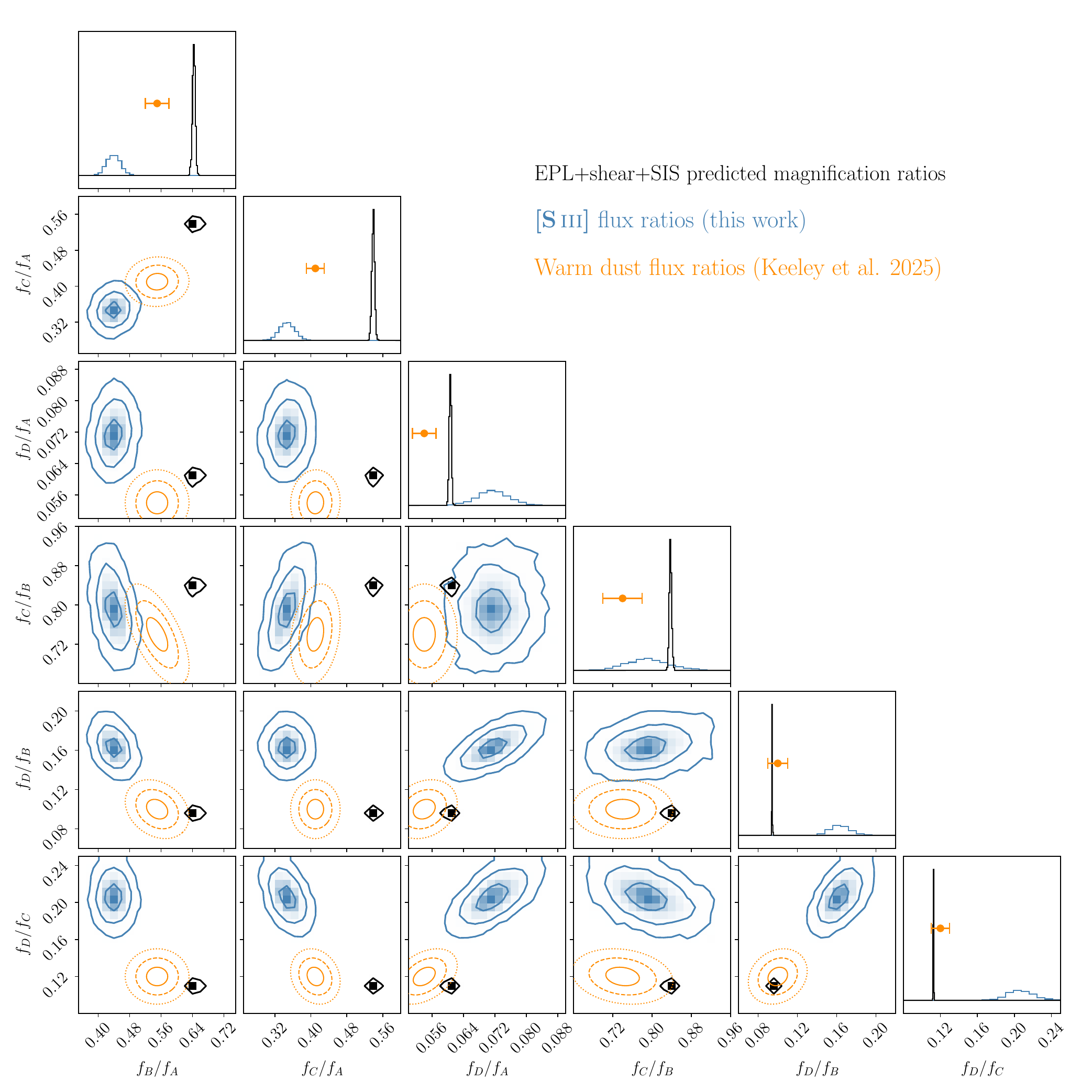}
    \caption{Joint posterior probability distribution of the ratios reported in Table~\ref{tab:results_FR}, for the [\ion{S}{III}] flux ratios (blue) and the magnification ratios predicted by the EPL+shear+SIS lens model (black). We also include the joint posterior on the warm dust flux ratios (orange) from \protect\cite{JWST_LQ_DM_III}. For the blue and orange posteriors, the contours correspond to 39\%, 86\%, and 99\% confidence regions (i.e., 1,2, and 3$\sigma$ for a 2D Gaussian distribution). For better visualization and legibility, the black posterior only shows the 99\% confidence, and contours have been slightly smoothed.}
    \vspace{-0.3cm}
    \label{fig:full_cov_FRs}
\end{figure*}

\begin{table*}
\caption{Pearson correlation coefficients between pairs of ratios reported in Table~\ref{tab:results_FR}, for the [\ion{S}{III}] flux ratios (upper left corner of each entry) and for the magnification ratios predicted by the EPL+shear+SIS lens model (lower right corner).}
\label{tab:cov_matrix_frs}
\centering
\begin{tabular}{c|c|c|c|c|c|c} 
\hline
 \diagbox[dir=NE]{$f_i/f_j$}{$\mu_i/\mu_j$} & $B/A$ & $C/A$ & $D/A$ & $C/B$ & $D/B$ & $D/C$ \\
  \hline 
 $B/A$ & 1 & \diagbox[dir=NE]{\phantom{$-$}0.24}{\phantom{$-$}0.87} & \diagbox[dir=NE]{\phantom{$-$}0.19}{\phantom{$-$}0.92} & \diagbox[dir=NE]{$-$0.50}{$-$0.23} & \diagbox[dir=NE]{$-$0.45}{$-$0.69} & \diagbox[dir=NE]{$-$0.02}{$-$0.44} \\
  \hline 
 $C/A$ & \diagbox[dir=NE]{\phantom{$-$}0.24}{\phantom{$-$}0.87} & 1 & \diagbox[dir=NE]{\phantom{$-$}0.18}{\phantom{$-$}0.94} & \diagbox[dir=NE]{\phantom{$-$}0.72}{\phantom{$-$}0.28} & \diagbox[dir=NE]{\phantom{$-$}0.01}{$-$0.35} & \diagbox[dir=NE]{$-$0.57}{$-$0.74}\\
  \hline 
 $D/A$ &  \diagbox[dir=NE]{\phantom{$-$}0.19}{\phantom{$-$}0.92} &\diagbox[dir=NE]{\phantom{$-$}0.18}{\phantom{$-$}0.94} & 1 & \diagbox[dir=NE]{\phantom{$-$}0.03}{\phantom{$-$}0.06} & \diagbox[dir=NE]{\phantom{$-$}0.79}{$-$0.36} & \diagbox[dir=NE]{\phantom{$-$}0.71}{$-$0.47 } \\
  \hline 
 $C/B$ & \diagbox[dir=NE]{$-$0.50}{$-$0.23} & \diagbox[dir=NE]{\phantom{$-$}0.72}{\phantom{$-$}0.28} & \diagbox[dir=NE]{\phantom{$-$}0.03}{\phantom{$-$}0.06} & 1 & \diagbox[dir=NE]{\phantom{$-$}0.34}{\phantom{$-$}0.67} & \diagbox[dir=NE]{$-$0.49}{$-$0.60} \\
  \hline 
  $D/B$ & \diagbox[dir=NE]{$-$0.45}{$-$0.69} & \diagbox[dir=NE]{\phantom{$-$}0.01}{$-$0.35} & \diagbox[dir=NE]{\phantom{$-$}0.79}{$-$0.36} & \diagbox[dir=NE]{\phantom{$-$}0.34}{\phantom{$-$}0.67} & 1& \diagbox[dir=NE]{\phantom{$-$}0.65}{\phantom{$-$}0.20} \\
   \hline 
   $D/C$ & \diagbox[dir=NE]{$-$0.02}{$-$0.44} & \diagbox[dir=NE]{$-$0.57}{$-$0.74} & \diagbox[dir=NE]{\phantom{$-$}0.71}{$-$0.47} & \diagbox[dir=NE]{$-$0.49}{$-$0.60} & \diagbox[dir=NE]{\phantom{$-$}0.65}{\phantom{$-$}0.20} & 1\\
  \hline  \hline
\end{tabular}
\vspace{-0.5cm}
\end{table*}

Even if we consider that the individual image fluxes $f_A$, $f_B$, $f_C$, $f_D$ (and, likewise, the magnifications $\mu_i$ (where $i = A,B,C,D$) predicted by the lens model at the quasar image positions) can be treated as uncorrelated quantities, the flux ratios formed from them are not. This is an instance of the spurious correlation of ratios first identified by \cite{Pearson1897}: in our case, ratios normalized to the same reference image (e.g., $f_B/f_A$ and $f_C/f_A$) both depend on the measured flux of image A, so a statistical fluctuation in $f_A$ shifts both ratios in the same direction simultaneously, inducing a positive correlation between them. The magnitude of the correlation between flux ratios can be approximated as \citep{Pearson1897, Yule1910}
\begin{equation}
    \rho\left(\frac{f_i}{f_k}, \frac{f_j}{f_l}\right) = \frac{r_{ij} v_i v_j + r_{kl} v_k v_l - r_{il} v_i v_l - r_{jk} v_j v_k}{\sqrt{v_i^2+v_k^2-2r_{ik} v_i v_k}\sqrt{v_j^2+v_l^2-2r_{jl} v_j v_l}}
    \label{eq:approx_pcc_ratio}
\end{equation}
where  $\rho$ is the Pearson correlation coefficient (PCC) between variables $f_i/f_k$ and $f_j/f_l$, $v_i$ is the coefficient of variation for flux $f_i$ and $r_{ij}$ is the PCC between $f_i$ and $f_j$. In particular, even if the flux values are independent (i.e., $r_{ij}=0$ for $i\neq j$), flux ratios sharing an image in common will be correlated \citep[e.g., Appendix C of][]{CAB_paper_I}. In general, however, flux/magnification values can be correlated due to the global lens model (for instance, the magnifications of the cusp images will be strongly correlated).

In Section~\ref{subsec:FR_values} (in particular, Table~\ref{tab:results_FR} and Figure~\ref{fig:FR_comparison}), we ignored correlations between flux ratios in order to facilitate the comparison with other emission feature and with results found in the literature. Correlations must be accounted for, however, when computing the flux-ratio likelihood in the forward modeling framework used for substructure inference \citep[e.g.,][]{JWST_LQ_DM_IV}. We therefore computed a full covariance matrix from the MCMC samples for the six ratios reported in Table~\ref{tab:results_FR}; for the measured [\ion{S}{III}] flux ratios, and for the magnification ratios predicted by the EPL+shear+SIS lens model from Section~\ref{subsubsec:lens_model}. The corresponding PCCs are reported in Table~\ref{tab:cov_matrix_frs}, and the full joint posteriors are displayed in Figure~\ref{fig:full_cov_FRs}, along with the joint posterior for the warm dust ratios from \cite{JWST_LQ_DM_III}. We find similar results (within one decimal place) when approximating the correlation coefficients with Equation~(\ref{eq:approx_pcc_ratio}).

\label{lastpage}
\end{document}